\documentclass[a4paper,aps,prd,10pt,preprintnumbers,showpacs,showkeys,twocolumn,superscriptaddress,nofootinbib,amsmath,amssymb]{revtex4-1}
\usepackage{cmap}
\usepackage[utf8]{inputenc}
\usepackage[T1]{fontenc}
\usepackage{graphicx,orcidlink,xcolor}

\def\Order#1{{\cal O}\left(#1\right)}

\begin{document}

\title{Long-lived quasinormal modes and echoes  in the Einstein-Gauss-Bonnet-Proca theory}

\author{B. C. Lütfüoğlu}
\email{bekir.lutfuoglu@uhk.cz}
\affiliation{Department of Physics, Faculty of Science, University of Hradec Kralove, \\
Rokitanskeho 62/26, Hradec Kralove, 500 03, Czech Republic.}

\begin{abstract}
We study quasinormal modes and time-domain profiles of a massive scalar field in the background of black holes arising in Einstein--Gauss--Bonnet--Proca gravity. The black holes in this theory possess \emph{primary Proca hair}, which modifies the effective potential and gives rise to distinctive dynamical phenomena. Using three complementary numerical techniques --- the WKB method with Padé resummation, time-domain integration, and the Frobenius (Leaver) method --- we obtain accurate spectra of quasinormal frequencies. Our analysis shows that increasing the scalar-field mass leads to arbitrarily long-lived states, known as quasi-resonances, a behavior shared by both the fundamental mode and the first overtone. The real part of the quasinormal mode decreases for the first and higher overtones and starting from the second overtone it reaches zero at some critical value of mass of the field.  When the effective potential develops an additional peak, the late-time signal exhibits a sequence of echoes.  Furthermore, the time-domain evolution reveals distinct regimes of intermediate power-law tails $\sim t^{-4/3-\ell}\sin(A(\mu)t)$ and a universal asymptotic tails $\sim t^{-5/6}\sin(\mu t)$. 
\end{abstract}

\keywords{black holes with primary hair; modified gravity; quasinormal modes; WKB method}

\maketitle

\section{Introduction}

The study of quasinormal modes (QNMs) has become an indispensable tool for understanding the dynamics of perturbed black holes. These damped oscillations, characterized by complex frequencies, carry unique signatures of the background geometry and provide a natural arena for testing gravitational theories in the strong-field regime. With the advent of gravitational-wave astronomy, increasingly precise measurements of ringdown signals have turned QNMs into a sensitive probe of possible deviations from general relativity~\cite{Berti:2009kk,Konoplya:2011qq,Cardoso:2016ryw,Bolokhov:2025uxz}.

An especially intriguing direction is the analysis of QNMs of \emph{massive} fields. Once a mass term is introduced, the phenomenology of black hole perturbations changes qualitatively. Massive fields may support arbitrarily long-lived modes, known as quasi-resonances~\cite{Ohashi:2004wr, Konoplya:2004wg}, and their late-time tails differ from the familiar monotonic power-law falloff of massless perturbations \cite{Konoplya:2004wg, Ohashi:2004wr, Ponglertsakul:2020ufm,  Gonzalez:2022upu, Burikham:2017gdm, 2753763,  Deng:2025uvp, Becar:2025niq, Chen:2025sbz,   Zhang:2018jgj, Aragon:2020teq}. Instead, oscillatory decay laws appear~\cite{Jing:2004zb, Koyama:2001qw, Moderski:2001tk, Rogatko:2007zz, Koyama:2001ee, Koyama:2000hj, Gibbons:2008gg, Gibbons:2008rs}, enriching the dynamical picture. This behavior has been observed for fields of various spin~\cite{Konoplya:2005hr, Fernandes:2021qvr, Konoplya:2017tvu, Percival:2020skc} and across diverse geometries, from standard black holes~\cite{Zhidenko:2006rs, Zinhailo:2018ska, Konoplya:2019hml, Bolokhov:2023bwm} to wormholes and other exotic compact objects~\cite{Churilova:2019qph}. Effective masses may also emerge in realistic settings, for instance in braneworld scenarios~\cite{Seahra:2004fg} or in the presence of magnetic fields~\cite{Konoplya:2007yy, Konoplya:2008hj, Wu:2015fwa, Kokkotas:2010zd}. Moreover, the prospect of a massive graviton, either as a fundamental particle or an effective description, links massive-field QNMs to current efforts such as Pulsar Timing Array searches for very long-wavelength gravitational waves~\cite{Konoplya:2023fmh, NANOGrav:2023hvm}. Nevertheless, quasi-resonances are not guaranteed: examples exist where massive fields fail to support infinitely long-lived modes~\cite{Zinhailo:2024jzt, Konoplya:2005hr}, underlining the importance of case-by-case analysis.

Parallel to these motivations, interest has grown in theories where black holes carry non-trivial \emph{primary hair}. In contrast to the usual situations where echoes arise only when additional structures are introduced externally—such as environmental matter distributions, quantum clouds near the horizon, or wormhole-like geometries~\cite{Cardoso:2016rao,Cardoso:2016oxy,Abedi:2016hgu,Mark:2017dnq,Bueno:2017hyj,Konoplya:2018yrp,Cardoso:2019apo,Bronnikov:2019sbx,Dong:2020odp,Churilova:2021tgn}—certain modified gravity models admit exact black hole solutions with intrinsic hair that can by itself produce echoes. A notable example was recently found in~\cite{Charmousis:2025jpx}, where scalar-tensor and vector-tensor interactions of Gauss–Bonnet type yield spherically symmetric black holes sourced by a Proca field. The corresponding action reads
\begin{equation}\label{action}
S = \int d^4x \sqrt{-g} \left( R - \alpha\, \mathcal{L}^{\text{VT}}_G - \beta\, \mathcal{L}^{\text{ST}}_G \right),
\end{equation}
with the vector-tensor and scalar-tensor terms given by
\begin{equation}
    \mathcal{L}_\mathcal{G}^{\rm VT} = 4 G^{\mu \nu}W_\mu W_\nu + 8 (W_\nu W^\nu) \nabla_\mu W^\mu + 6(W_\nu W^\nu)^2,
\end{equation}
\begin{equation}
    \mathcal{L}_\mathcal{G}^{\rm ST} = \phi \mathcal{G} - 4 G^{\mu \nu} \nabla_\mu \phi \nabla_\nu \phi - 4 \Box \phi (\partial \phi)^2 - 2 (\partial \phi)^4,
\end{equation}
where $\phi$ is a scalar field, $W_\mu$ is a Proca vector, and
\begin{equation}
    \mathcal{G} = R^2 - 4 R_{\mu \nu} R^{\mu \nu} + R_{\mu \nu \alpha \beta} R^{\mu \nu \alpha \beta}, \quad 
    G_{\mu \nu} = R_{\mu \nu} - \tfrac{1}{2}g_{\mu\nu}R,
\end{equation}
denote the Gauss–Bonnet invariant and Einstein tensor, respectively. Depending on the relation between $\alpha$ and $\beta$, the resulting metrics admit either rational or square-root forms, both asymptotically flat and reducing to Schwarzschild in the appropriate limit. It is worth of mentioning that the literature on perturbations and quasinormal modes of Einstein-Gauss-Bonnet black holes without additional fields, such as Proca, is considerable and here we will mention only to a few works \cite{Konoplya:2020cbv,Cuyubamba:2020moe,Konoplya:2020bxa}.

These spacetimes offer a clean setup for testing the role of Proca hair. Their geodesics, shadows, and grey-body factors have been analyzed in~\cite{Lutfuoglu:2025ldc}, while the QNM spectrum of massless fields and the associated appearance of echoes were studied in~\cite{Konoplya:2025uiq}. The latter work highlighted that primary hair alone is sufficient to trigger strong echoes, without any external modifications of the effective potential.

The present paper extends this program to \emph{massive} fields in the Einstein–Proca–Gauss–Bonnet background. As argued in~\cite{Konoplya:2024wds}, echoes become even more pronounced when the perturbing field carries mass, since the slow decay of massive modes amplifies their visibility in the late-time signal. Here we demonstrate that the spectrum indeed accommodates quasi-resonances and that, for certain ranges of parameters, echoes manifest as distinctive modifications of the oscillatory tails of massive perturbations.

The remainder of this work is organized as follows. In Sec.~II we review the background geometry and presents the perturbation equations for a massive scalar field. Sec.~III describes all three numerical methods employed in both frequency and time domains: time-domain integration, Frobenius method and WKB approach. Our results for quasinormal modes, quasi-resonances, and echo signals are summarized in Sec.~IV. Finally, Sec.~V contains our conclusions and outlook.

\section{The black hole metric and wave-like equation}\label{sec:background}

We now turn to the wave-like equation for the static, spherically symmetric black hole geometries obtained in~\cite{Charmousis:2025jpx}, which follow from the action~\eqref{action} and include both vector-tensor and scalar-tensor Gauss--Bonnet couplings. Their common line element reads
\begin{equation}
ds^2 = -f(r)\, dt^2 + \frac{dr^2}{f(r)} + r^2 (d\theta^2 + \sin^2 \theta\, d\phi^2),
\end{equation}
where the function \( f(r) \) is determined by the values of the couplings \( \alpha \) and \( \beta \).

For generic values, \( \alpha \neq \beta \), the metric function takes the form \cite{Charmousis:2025jpx}
\begin{align}\label{metricfunc}
f(r) &= 1 - \frac{2 \alpha (M - Q)}{r (\alpha + \beta)} + \frac{r^2}{2 (\alpha + \beta)} \\\nonumber
& - \frac{r^2}{2 (\alpha + \beta)}
\sqrt{1 + \frac{8 \alpha Q}{r^3} + \frac{8 \beta M}{r^3} - \frac{16 \alpha \beta (M - Q)^2}{r^6} },
\end{align}
with \( M \) the ADM mass and \( Q \) an independent constant linked to the Proca field. Since \( Q \) is not constrained by \( M \), the solutions carry \emph{primary hair}, i.e. charges not reducible to standard conserved quantities. When the relation \( Q = M \) holds, the spacetime reduces to the black holes of four-dimensional scalar--tensor Einstein--Gauss--Bonnet gravity~\cite{Lu:2020iav,Kobayashi:2020wqy,Fernandes:2020nbq}.

A special simplification occurs in the limit \( \beta \to -\alpha \), where the expression~\eqref{metricfunc} collapses to a rational form:
\begin{align}\label{simplified}
\lim_{\beta\to-\alpha}f(r) &= \frac{r^3}{r^3 - 4 \alpha (M - Q)} \Bigg( 1 - \frac{2 M}{r}
 \\\nonumber
&\qquad  - \frac{4 \alpha (M - Q)}{r^3} + \frac{4 \alpha (M - Q)^2}{r^4} \Bigg).
\end{align}
The geometry remains regular as long as the denominator does not vanish outside the horizon. In both families of solutions, Schwarzschild spacetime is recovered when the couplings vanish, \( \alpha = \beta = 0 \).

The parameter space admitting black holes has been analyzed in detail in~\cite{Charmousis:2025jpx}. It was shown that large positive values of \( \alpha \) and \( \beta \) may prevent horizon formation if \( \alpha + \beta \neq 0 \), while for the case \( \alpha + \beta = 0 \) the same obstruction arises for sufficiently large values of \( Q \). In such cases, the solutions correspond instead to naked singularities or horizonless compact objects.

To explore the quasinormal modes of these backgrounds, we consider test fields propagating on the fixed metric. In what follows, we focus on massless scalar and Dirac fields. In both cases, the dynamics can be reduced to a Schrödinger-type master equation:
\begin{equation}\label{master_eq}
\frac{d^2 \Psi}{dr_*^2} - \left[ V(r) -  \omega^2\right] \Psi = 0,
\end{equation}
where \( \omega \) is the complex quasinormal frequency and the tortoise coordinate \( r_* \) is defined as
\begin{equation}
\frac{dr_*}{dr} = f(r)^{-1}.
\end{equation}

For scalar perturbations, the effective potential is
\begin{equation}
V(r) = f(r) \left( \frac{\ell(\ell+1)}{r^2} + \frac{f'(r)}{r} +\mu^2 \right),
\end{equation}
with multipole number \(\ell = 0,1,2,\dots\).

\begin{figure}
\resizebox{\linewidth}{!}{\includegraphics{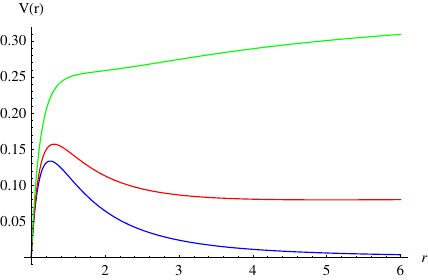}}
\caption{Effective potentials for $\ell=0$, $\alpha=-\beta=0.1$, $Q=0.1$, $r_{h}=1$, $\mu=0$ (blue, bottom), $\mu=0.3$ (red, middle) and $\mu=0.6$ (green, top). }\label{fig:potentials1}
\end{figure}

From Fig.~\ref{fig:potentials1}, we see that the effective potential approaches $\mu^2$ at large distances. Moreover, when the field mass exceeds a certain critical value (which depends on $\ell$), the potential no longer develops a peak.

\section{Three methods for calculations of quasinormal modes}\label{sec:numerical}

To explore the quasinormal spectrum we rely on the three well-established but complementary techniques: Frobenius or Leaver method, direct evolution in the time domain and semi-analytical estimates obtained from the WKB method with Padé resummation. 

The QNM boundary conditions imply the following behavior 
\begin{equation}
\Psi(r) \propto
\begin{cases}
e^{- i \omega r_{*}}, & r \to r_{+} ,\\[4pt] e^{+ i \Omega r_{*}}, & r \to \infty,
\end{cases}
\label{eq:BCs}
\end{equation}
where
\begin{equation}
\Omega = \sqrt{\omega^{2} - \mu^{2}}.
\end{equation}
The square root is chosen in such a way that $\mathrm{Re}(\Omega)$ and $\mathrm{Re}(\omega)$ have the same sign \cite{Zhidenko:2006rs}.

\textbf{Time-domain integration.} For the full dynamical profile we employ the characteristic algorithm of Gundlach–Price–Pullin~\cite{Gundlach:1993tp}. In terms of null coordinates \(u=t-r_*\), \(v=t+r_*\), the master equation~\eqref{master_eq} becomes
\[
4\frac{\partial^2 \Psi}{\partial u \partial v} + V(r_*)\Psi = 0,
\]
which is discretized on a two-dimensional grid. The evolution at a new point \(N\) is obtained from its neighbors via
\begin{eqnarray}\label{Discretization}
\Psi(N) &=& \Psi(W)+\Psi(E)-\Psi(S) \\
&&\nonumber - \, h^2 V(S)\,\frac{\Psi(W)+\Psi(E)}{8}+\Order{h^4}.
\end{eqnarray}
Here \(N=(u+h,v+h)\), \(W=(u+h,v)\), \(E=(u,v+h)\), and \(S=(u,v)\).  

Initial data are specified on the null lines. Gaussian pulses centered near the potential maximum are chosen so that the transient dies out quickly, leaving the clean ringdown signal~\cite{Konoplya:2011qq}. The resulting profile naturally exhibits three phases: an initial outburst, a stage of exponentially damped oscillations, and power-law late-time tails.  

From the numerical waveform one can extract the frequencies by fitting the signal to a sum of damped exponentials using the Prony method,
\[
\Psi(t) = \sum_{k=1}^N A_k e^{-i\omega_k t}.
\]
In this way the dominant quasinormal frequencies can be determined directly from the evolution, without assuming any particular approximation for the potential. The time-domain integration method has been effectively used for finding the dominant quasinormal modes and testing the (in)stability of black holes and wormholes \cite{Zhidenko:2008fp,Konoplya:2020jgt,Lutfuoglu:2025hwh,Lutfuoglu:2025hjy,Konoplya:2020juj,Skvortsova:2024atk,Skvortsova:2023zmj,Malik:2024qsz}.

\textbf{WKB method.} While time-domain data give a robust picture of the evolution, the WKB method offers a faster semi-analytic estimate of the frequencies. This technique, originally developed in~\cite{Schutz:1985km,Iyer:1986np,Konoplya:2003ii}, is particularly efficient when the effective potential forms a single barrier. The method consists of matching approximate solutions across the peak of the potential and yields the quantization rule
\begin{equation}
\frac{i (\omega^2 - V_0)}{\sqrt{-2 V_0''}} - \sum_{j=2}^{k} \Lambda_j = n + \tfrac{1}{2}, 
\quad n=0,1,2,\ldots,
\end{equation}
where \(V_0\) is the height of the potential barrier, \(V_0''\) its second derivative at the maximum, and \(\Lambda_j\) are higher-order correction terms. The coefficients \(\Lambda_j\) are known up to the 13th order~\cite{Matyjasek:2017psv}.

Although the WKB expansion is asymptotic rather than convergent, its accuracy can be improved by Padé resummation~\cite{Konoplya:2019hlu}. The idea is to reorganize the truncated series as a rational function in the small parameter \(\epsilon\),
\begin{equation}
P_{\tilde{n}/\tilde{m}}=
\frac{Q_0+Q_1\epsilon+\ldots+Q_{\tilde{n}}\epsilon^{\tilde{n}}}
{R_0+R_1\epsilon+\ldots+R_{\tilde{m}}\epsilon^{\tilde{m}}},
\end{equation}
with \(\tilde{n}+\tilde{m}=k\). Balanced approximants, e.g. \(P_{6/6}\) for order 12 or \(P_{6/7}\) for order 13, usually deliver the best results.  

In practice, the method performs very well for the fundamental modes (see, \cite{Lutfuoglu:2025ljm,Konoplya:2023ahd,Dubinsky:2024aeu,Dubinsky:2024hmn,Dubinsky:2025fwv,Konoplya:2005sy,Bolokhov:2025lnt,Bolokhov:2024ixe,Malik:2024nhy,Konoplya:2024lch,Skvortsova:2024wly} for applications). Here we used the WKB method for finding an initial guess which was further used in the Leaver (Frobenius) method.

\begin{figure*}
\resizebox{\linewidth}{!}{\includegraphics{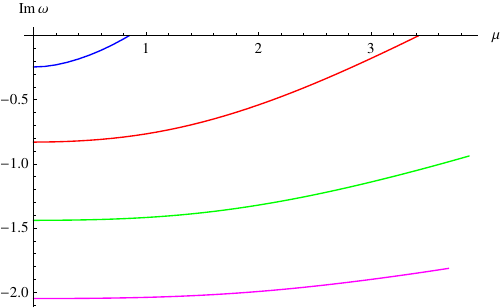}\includegraphics{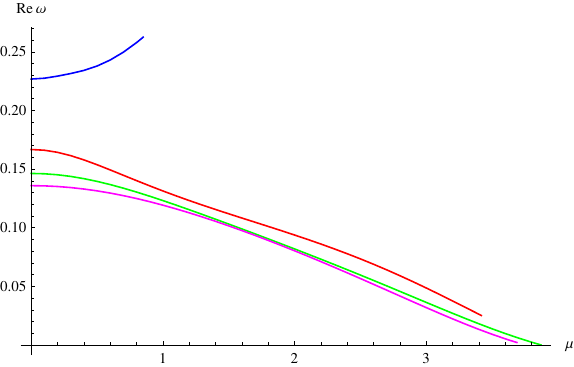}}
\caption{Quasinormal modes for the fundamental mode and first three overtones at $r_{h}=1$, $Q=0.1$, $\alpha=-\beta=0.1$ found by the precise Leaver method.}\label{fig:Frobenius}
\end{figure*}
\begin{figure}
\resizebox{\linewidth}{!}{\includegraphics{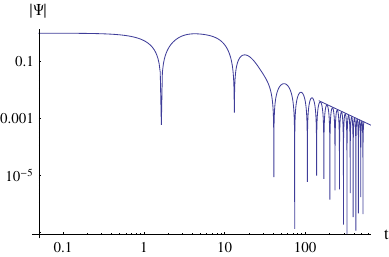}}
\caption{Time-domain profile for $\ell=0$, $r_{h}=1$, $Q=0.1$, $\alpha=-\beta =0.1$, $\mu=0.1$. The intermediate envelope for the tails $\approx t^{-4/3}$. The first two oscillations represent period of dominance of quasinormal modes. The Prony method gives $\omega = 0.23 - 0.24 i$, while the precise Frobenius method gives $\omega =0.227367 - 0.2412166 i$.}\label{fig:TD1}
\end{figure}
\begin{figure*}
\resizebox{\linewidth}{!}{\includegraphics{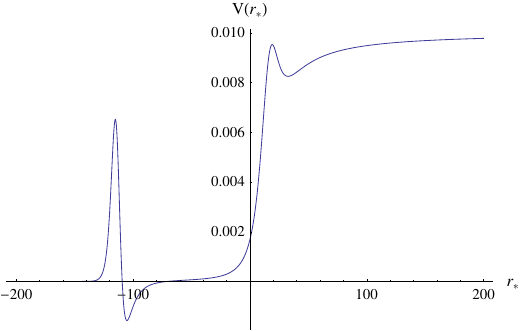}\includegraphics{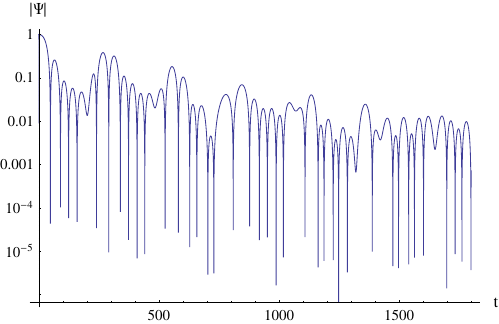}}
\caption{Effective potential (left) and time-domain profile (right) for $\ell=0$, $r_{h}=1$, $Q=1.16366$, $\alpha=-\beta =-4.13757$, $\mu=0.1$.}\label{fig:TD2}
\end{figure*}
\begin{figure*}
\resizebox{\linewidth}{!}{\includegraphics{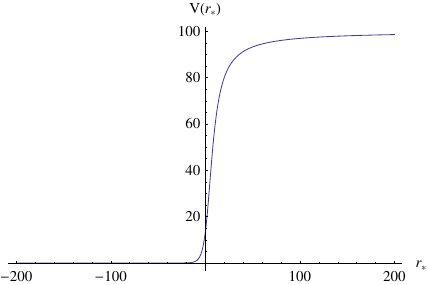}\includegraphics{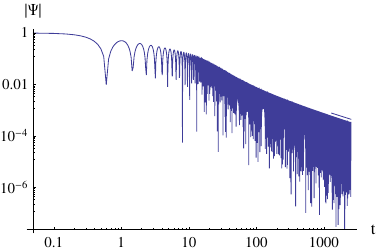}}
\caption{Effective potential (left) and time-domain profile (right) for $\ell=0$, $r_{h}=1$, $Q=1.7$, $\alpha=-\beta =0.1$, $\mu=10$.}\label{fig:TD3}
\end{figure*}
\begin{figure}
\resizebox{\linewidth}{!}{\includegraphics{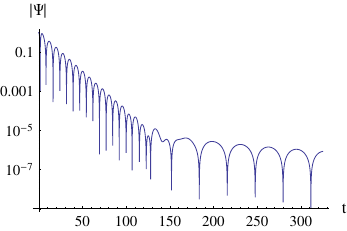}}
\caption{Time-domain profile for $\ell=1$, $r_{h}=1$, $Q=0.5$, $\alpha=0.5$, $\beta =0.5$, $\mu=0.1$. The Prony method gives the $\omega = 0.4158 - 0.0918 i$. The WKB method gives $\omega = 0.415645 - 0.0917889 i$.}\label{fig:TD4}
\end{figure}

\textbf{Leaver method.}  Among the available numerical approaches, the Frobenius expansion (often called the Leaver method~\cite{Leaver:1985ax}) remains one of the most accurate techniques for computing black hole quasinormal modes. Its key idea is to factor out the singular behavior of the solution at the physical boundaries and expand the remaining part of the wave function into a power series.

To implement this, the radial wave function is written as
\begin{equation}
\Psi(r) = A(r,\omega)\, y(r),
\label{eq:factorization}
\end{equation}
where the prefactor \(A(r,\omega)\) encodes the near-horizon ingoing behavior and the correct asymptotic falloff, leaving \(y(r)\) analytic both at the event horizon and at spatial infinity. The regular function \(y(r)\) is then expanded in a Frobenius series around the horizon,
\begin{equation}
y(r) = \sum_{k=0}^\infty a_k \left( \frac{r-r_+}{r} \right)^k,
\label{eq:Frobenius_series}
\end{equation}
which converges for the entire exterior region \(r \geq r_+\). In this way the boundary condition at the horizon is automatically enforced, while the quasinormal spectrum arises from the requirement of convergence at infinity.

Substituting the expansion~\eqref{eq:Frobenius_series} into the radial equation produces a recurrence relation for the coefficients,
\begin{equation}
\alpha_n a_{n+1} + \beta_n a_n + \gamma_n a_{n-1} = 0,
\quad n \geq 0, 
\quad \gamma_0 = 0,
\label{eq:recurrence_relation}
\end{equation}
with \(\alpha_n\), \(\beta_n\), \(\gamma_n\) depending on the frequency and black hole parameters. The coefficients \(\alpha_n\), \(\beta_n\), and \(\gamma_n\) can be calculated numerically via Gaussian eliminations from the initial 15-terms recurrence relation.

The quasinormal mode condition follows from requiring the series to converge at infinity. This translates into a continued-fraction equation,
\begin{equation}
\beta_{n} - \frac{\alpha_{n-1} \gamma_{n}}{\beta_{n-1} -}
\frac{\alpha_{n-2} \gamma_{n-1}}{\beta_{n-2} - \cdots}
= \frac{\alpha_{n} \gamma_{n+1}}{\beta_{n+1} -}
\frac{\alpha_{n+1} \gamma_{n+2}}{\beta_{n+2} - \cdots},
\label{eq:continued_fraction}
\end{equation}
which is solved numerically by truncating the infinite fraction at some large order \(n_{\mathrm{max}}\). The roots of this equation give the precise quasinormal frequencies.

The convergence of the fraction can become slow for modes with large damping. In such situations, the Nollert improvement~\cite{Nollert:1993zz} is employed. It makes use of the asymptotic expansion of the coefficient ratio
\begin{equation}
R_n \equiv \frac{a_{n+1}}{a_n} = C_0 + \frac{C_1}{\sqrt{n}} + \frac{C_2}{n} + \cdots,
\label{eq:Nollert_asymptotics}
\end{equation}
to approximate the tail of the series and replace the truncated part of the continued fraction. This significantly accelerates convergence and enhances numerical stability.

In practice, the Frobenius (Leaver) method combined with the Nollert technique provides one of the most reliable tools for computing the spectrum. It allows high-precision determination of both the fundamental frequencies and higher overtones, and is applicable to a wide range of perturbations, including massless and massive scalar fields, Dirac fields, and charged backgrounds, once the correct asymptotic conditions are imposed~\cite{Konoplya:2004uk, Dias:2022oqm, Stuchlik:2025mjj, Xiong:2023usm, Kanti:2006ua, Zinhailo:2024kbq, Konoplya:2007zx, Stuchlik:2025ezz}.

\section{Quasinormal modes and echoes}\label{sec:qnms}

In fact, the general form of the metric is given by eq.~\ref{metricfunc}, while the special case $\alpha = -\beta$ is described by eq.~\ref{simplified}. Although our conclusions apply to both configurations, in the frequency-domain analysis we focus on the latter case, since its metric function and effective potential take a rational form, allowing the precise Leaver method to be applied directly.

We observe that the fundamental mode and the first overtone exhibit qualitatively different behavior compared to the second and, apparently, higher overtones (see Fig.~\ref{fig:Frobenius}). As the field mass $\mu$ increases, the damping rate of the fundamental mode monotonically decreases and eventually approaches zero at some critical value of $\mu$. At this point, the fundamental mode disappears from the spectrum, and the first overtone takes its place as the new fundamental mode. When $\mu$ grows further, the first overtone undergoes a similar transition: its damping rate decreases to zero, and it also disappears, so that the second overtone becomes dominant.  

The behavior of the second overtone, however, differs in an essential way. For this mode, both the damping rate and the real oscillation frequency decrease with increasing $\mu$, but the real part of the frequency approaches zero faster than the damping rate. Consequently, the second overtone disappears from the spectrum when $\mathrm{Re}(\omega)\to 0$, before its damping rate fully vanishes. Higher overtones appear to follow the same pattern. This type of behavior has also been observed recently in certain brane-world models~\cite{Zinhailo:2024jzt}, suggesting that it may be a more general feature of theories with additional scales.  

In the time domain, we identify two qualitatively distinct regimes of evolution, illustrated in Figs.~\ref{fig:TD1}--\ref{fig:TD4}. For parameter ranges where the effective potential has a single smooth peak, the signal decays through the familiar oscillatory power-law tails, similar to the Schwarzschild case. The intermediate tails are also power-law, but differ from the asymptotic ones. Specifically, for $\ell=0$ we find an intermediate decay of the form 
\[
|\Psi| \sim t^{-4/3},
\]
while at asymptotically late times the decay coincides with the well-known Schwarzschild and Reissner–Nordström results~\cite{Koyama:2000hj,Koyama:2001ee},
\begin{equation}
|\Psi| \sim t^{-5/6} \sin(\mu t), 
\qquad t \to \infty,
\tag{19}
\end{equation}
dominating once 
\begin{equation}
\frac{t}{M} \gg (\mu M)^{-3}.
\tag{20}
\end{equation}
The intermediate decay law can be written more generally as 
\begin{equation}
|\Psi| \sim t^{-\frac{4}{3}-\ell} \sin\!\big(A(\mu)t\big),
\tag{22}
\end{equation}
and this coincides with the behavior reported for the Bardeen spacetime~\cite{Bolokhov:2023ruj}.  

A qualitatively different picture emerges when the effective potential develops a second peak, as discussed in~\cite{Konoplya:2025uiq}. Fig.~\ref{fig:TD2} shows an example of such a potential and the corresponding late-time evolution of a massive field. In this case, the signal exhibits a sequence of \emph{echoes}, superimposed on the oscillatory asymptotic tails. A notable feature of this regime is the very slow decay of echoes: even at extremely late times their amplitude remains only one to two orders of magnitude smaller than that of the initial ringdown stage. This persistence makes echoes an especially distinctive signature of the Proca–Gauss–Bonnet background, potentially relevant for gravitational-wave phenomenology.

\section{Conclusions}\label{sec:conclusions}
In this work we have analyzed the quasinormal spectra and time-domain evolution of a massive scalar field in the background of Einstein--Gauss--Bonnet--Proca black holes. Our main results can be summarized as follows:

\begin{enumerate}
\item The quasinormal spectrum exhibits quasi-resonances: as the field mass increases, the damping rate of the fundamental mode approaches zero at a critical value of $\mu$, after which the fundamental mode disappears from the spectrum and is replaced by the first overtone. At larger $\mu$, a similar transition occurs for the first overtone, while higher modes demonstrate qualitatively different behavior, with their real oscillation frequencies tending to zero faster than the damping rates.
\item In the time domain, we identified two distinct regimes of late-time decay. For single-barrier effective potentials, the signal shows intermediate oscillatory tails with decay $\sim t^{-4/3-\ell}\sin(A(\mu)t)$, while at asymptotically late times the universal massive-field tail $\sim t^{-5/6}\sin(\mu t)$ dominates. These results are consistent with earlier analyses of massive fields in some other black hole spacetimes.
\item When the effective potential develops an additional peak, the ringdown signal contains a series of echoes. Importantly, as shown in \cite{Konoplya:2025uiq}  for massles fields these echoes arise \emph{solely from the presence of primary Proca hair}, without the need to invoke exotic compact objects or external matter distributions. Here we confirm this qualitatively new mechanism for echo production for a massive field.
\end{enumerate}

In this work, we did not attempt to provide an exhaustive analysis of the spectrum’s dependence on all parameters, since the behavior with respect to parameters other than the field mass $\mu$ closely resembles the massless case studied in~\cite{Konoplya:2025uiq}. Numerical data for specific values of the modes at fixed parameter choices are available from the author upon reasonable request.

\acknowledgments
The author is grateful to Excellence Project FoS UHK 2205/2025-2026 for the financial support.

\bibliography{bibliography}

\begin{thebibliography}{96}%
\makeatletter
\providecommand \@ifxundefined [1]{%
 \@ifx{#1\undefined}
}%
\providecommand \@ifnum [1]{%
 \ifnum #1\expandafter \@firstoftwo
 \else \expandafter \@secondoftwo
 \fi
}%
\providecommand \@ifx [1]{%
 \ifx #1\expandafter \@firstoftwo
 \else \expandafter \@secondoftwo
 \fi
}%
\providecommand \natexlab [1]{#1}%
\providecommand \enquote  [1]{``#1''}%
\providecommand \bibnamefont  [1]{#1}%
\providecommand \bibfnamefont [1]{#1}%
\providecommand \citenamefont [1]{#1}%
\providecommand \href@noop [0]{\@secondoftwo}%
\providecommand \href [0]{\begingroup \@sanitize@url \@href}%
\providecommand \@href[1]{\@@startlink{#1}\@@href}%
\providecommand \@@href[1]{\endgroup#1\@@endlink}%
\providecommand \@sanitize@url [0]{\catcode `\\12\catcode `\$12\catcode `\&12\catcode `\#12\catcode `\^12\catcode `\_12\catcode `\%12\relax}%
\providecommand \@@startlink[1]{}%
\providecommand \@@endlink[0]{}%
\providecommand \url  [0]{\begingroup\@sanitize@url \@url }%
\providecommand \@url [1]{\endgroup\@href {#1}{\urlprefix }}%
\providecommand \urlprefix  [0]{URL }%
\providecommand \Eprint [0]{\href }%
\providecommand \doibase [0]{http://dx.doi.org/}%
\providecommand \selectlanguage [0]{\@gobble}%
\providecommand \bibinfo  [0]{\@secondoftwo}%
\providecommand \bibfield  [0]{\@secondoftwo}%
\providecommand \translation [1]{[#1]}%
\providecommand \BibitemOpen [0]{}%
\providecommand \bibitemStop [0]{}%
\providecommand \bibitemNoStop [0]{.\EOS\space}%
\providecommand \EOS [0]{\spacefactor3000\relax}%
\providecommand \BibitemShut  [1]{\csname bibitem#1\endcsname}%
\let\auto@bib@innerbib\@empty
\bibitem [{\citenamefont {Berti}\ \emph {et~al.}(2009)\citenamefont {Berti}, \citenamefont {Cardoso},\ and\ \citenamefont {Starinets}}]{Berti:2009kk}%
  \BibitemOpen
  \bibfield  {author} {\bibinfo {author} {\bibfnamefont {E.}~\bibnamefont {Berti}}, \bibinfo {author} {\bibfnamefont {V.}~\bibnamefont {Cardoso}}, \ and\ \bibinfo {author} {\bibfnamefont {A.~O.}\ \bibnamefont {Starinets}},\ }\href {\doibase 10.1088/0264-9381/26/16/163001} {\bibfield  {journal} {\bibinfo  {journal} {Class. Quant. Grav.}\ }\textbf {\bibinfo {volume} {26}},\ \bibinfo {pages} {163001} (\bibinfo {year} {2009})},\ \Eprint {http://arxiv.org/abs/0905.2975} {arXiv:0905.2975 [gr-qc]} \BibitemShut {NoStop}%
\bibitem [{\citenamefont {Konoplya}\ and\ \citenamefont {Zhidenko}(2011)}]{Konoplya:2011qq}%
  \BibitemOpen
  \bibfield  {author} {\bibinfo {author} {\bibfnamefont {R.~A.}\ \bibnamefont {Konoplya}}\ and\ \bibinfo {author} {\bibfnamefont {A.}~\bibnamefont {Zhidenko}},\ }\href {\doibase 10.1103/RevModPhys.83.793} {\bibfield  {journal} {\bibinfo  {journal} {Rev. Mod. Phys.}\ }\textbf {\bibinfo {volume} {83}},\ \bibinfo {pages} {793} (\bibinfo {year} {2011})},\ \Eprint {http://arxiv.org/abs/1102.4014} {arXiv:1102.4014 [gr-qc]} \BibitemShut {NoStop}%
\bibitem [{\citenamefont {Cardoso}\ and\ \citenamefont {Gualtieri}(2016)}]{Cardoso:2016ryw}%
  \BibitemOpen
  \bibfield  {author} {\bibinfo {author} {\bibfnamefont {V.}~\bibnamefont {Cardoso}}\ and\ \bibinfo {author} {\bibfnamefont {L.}~\bibnamefont {Gualtieri}},\ }\href {\doibase 10.1088/0264-9381/33/17/174001} {\bibfield  {journal} {\bibinfo  {journal} {Class. Quant. Grav.}\ }\textbf {\bibinfo {volume} {33}},\ \bibinfo {pages} {174001} (\bibinfo {year} {2016})},\ \Eprint {http://arxiv.org/abs/1607.03133} {arXiv:1607.03133 [gr-qc]} \BibitemShut {NoStop}%
\bibitem [{\citenamefont {Bolokhov}\ and\ \citenamefont {Skvortsova}(2025{\natexlab{a}})}]{Bolokhov:2025uxz}%
  \BibitemOpen
  \bibfield  {author} {\bibinfo {author} {\bibfnamefont {S.~V.}\ \bibnamefont {Bolokhov}}\ and\ \bibinfo {author} {\bibfnamefont {M.}~\bibnamefont {Skvortsova}},\ }\href@noop {} {\  (\bibinfo {year} {2025}{\natexlab{a}})},\ \Eprint {http://arxiv.org/abs/2504.05014} {arXiv:2504.05014 [gr-qc]} \BibitemShut {NoStop}%
\bibitem [{\citenamefont {Ohashi}\ and\ \citenamefont {Sakagami}(2004)}]{Ohashi:2004wr}%
  \BibitemOpen
  \bibfield  {author} {\bibinfo {author} {\bibfnamefont {A.}~\bibnamefont {Ohashi}}\ and\ \bibinfo {author} {\bibfnamefont {M.-a.}\ \bibnamefont {Sakagami}},\ }\href {\doibase 10.1088/0264-9381/21/16/010} {\bibfield  {journal} {\bibinfo  {journal} {Class. Quant. Grav.}\ }\textbf {\bibinfo {volume} {21}},\ \bibinfo {pages} {3973} (\bibinfo {year} {2004})},\ \Eprint {http://arxiv.org/abs/gr-qc/0407009} {arXiv:gr-qc/0407009} \BibitemShut {NoStop}%
\bibitem [{\citenamefont {Konoplya}\ and\ \citenamefont {Zhidenko}(2005)}]{Konoplya:2004wg}%
  \BibitemOpen
  \bibfield  {author} {\bibinfo {author} {\bibfnamefont {R.~A.}\ \bibnamefont {Konoplya}}\ and\ \bibinfo {author} {\bibfnamefont {A.~V.}\ \bibnamefont {Zhidenko}},\ }\href {\doibase 10.1016/j.physletb.2005.01.078} {\bibfield  {journal} {\bibinfo  {journal} {Phys. Lett. B}\ }\textbf {\bibinfo {volume} {609}},\ \bibinfo {pages} {377} (\bibinfo {year} {2005})},\ \Eprint {http://arxiv.org/abs/gr-qc/0411059} {arXiv:gr-qc/0411059} \BibitemShut {NoStop}%
\bibitem [{\citenamefont {Ponglertsakul}\ and\ \citenamefont {Gwak}(2020)}]{Ponglertsakul:2020ufm}%
  \BibitemOpen
  \bibfield  {author} {\bibinfo {author} {\bibfnamefont {S.}~\bibnamefont {Ponglertsakul}}\ and\ \bibinfo {author} {\bibfnamefont {B.}~\bibnamefont {Gwak}},\ }\href {\doibase 10.1140/epjc/s10052-020-08616-1} {\bibfield  {journal} {\bibinfo  {journal} {Eur. Phys. J. C}\ }\textbf {\bibinfo {volume} {80}},\ \bibinfo {pages} {1023} (\bibinfo {year} {2020})},\ \Eprint {http://arxiv.org/abs/2007.16108} {arXiv:2007.16108 [gr-qc]} \BibitemShut {NoStop}%
\bibitem [{\citenamefont {Gonz\'alez}\ \emph {et~al.}(2022)\citenamefont {Gonz\'alez}, \citenamefont {Papantonopoulos}, \citenamefont {Saavedra},\ and\ \citenamefont {V\'asquez}}]{Gonzalez:2022upu}%
  \BibitemOpen
  \bibfield  {author} {\bibinfo {author} {\bibfnamefont {P.~A.}\ \bibnamefont {Gonz\'alez}}, \bibinfo {author} {\bibfnamefont {E.}~\bibnamefont {Papantonopoulos}}, \bibinfo {author} {\bibfnamefont {J.}~\bibnamefont {Saavedra}}, \ and\ \bibinfo {author} {\bibfnamefont {Y.}~\bibnamefont {V\'asquez}},\ }\href {\doibase 10.1007/JHEP06(2022)150} {\bibfield  {journal} {\bibinfo  {journal} {JHEP}\ }\textbf {\bibinfo {volume} {06}},\ \bibinfo {pages} {150} (\bibinfo {year} {2022})},\ \Eprint {http://arxiv.org/abs/2204.01570} {arXiv:2204.01570 [gr-qc]} \BibitemShut {NoStop}%
\bibitem [{\citenamefont {Burikham}\ \emph {et~al.}(2017)\citenamefont {Burikham}, \citenamefont {Ponglertsakul},\ and\ \citenamefont {Tannukij}}]{Burikham:2017gdm}%
  \BibitemOpen
  \bibfield  {author} {\bibinfo {author} {\bibfnamefont {P.}~\bibnamefont {Burikham}}, \bibinfo {author} {\bibfnamefont {S.}~\bibnamefont {Ponglertsakul}}, \ and\ \bibinfo {author} {\bibfnamefont {L.}~\bibnamefont {Tannukij}},\ }\href {\doibase 10.1103/PhysRevD.96.124001} {\bibfield  {journal} {\bibinfo  {journal} {Phys. Rev. D}\ }\textbf {\bibinfo {volume} {96}},\ \bibinfo {pages} {124001} (\bibinfo {year} {2017})},\ \Eprint {http://arxiv.org/abs/1709.02716} {arXiv:1709.02716 [gr-qc]} \BibitemShut {NoStop}%
\bibitem [{\citenamefont {Malik}(2024)}]{2753763}%
  \BibitemOpen
  \bibfield  {author} {\bibinfo {author} {\bibfnamefont {Z.}~\bibnamefont {Malik}},\ }\href {\doibase 10.13140/RG.2.2.32912.99849} {\  (\bibinfo {year} {2024}),\ 10.13140/RG.2.2.32912.99849}\BibitemShut {NoStop}%
\bibitem [{\citenamefont {Deng}\ \emph {et~al.}(2025)\citenamefont {Deng}, \citenamefont {Liu}, \citenamefont {Long}, \citenamefont {Xiao},\ and\ \citenamefont {Jing}}]{Deng:2025uvp}%
  \BibitemOpen
  \bibfield  {author} {\bibinfo {author} {\bibfnamefont {W.}~\bibnamefont {Deng}}, \bibinfo {author} {\bibfnamefont {W.}~\bibnamefont {Liu}}, \bibinfo {author} {\bibfnamefont {F.}~\bibnamefont {Long}}, \bibinfo {author} {\bibfnamefont {K.}~\bibnamefont {Xiao}}, \ and\ \bibinfo {author} {\bibfnamefont {J.}~\bibnamefont {Jing}},\ }\href@noop {} {\  (\bibinfo {year} {2025})},\ \Eprint {http://arxiv.org/abs/2507.13978} {arXiv:2507.13978 [gr-qc]} \BibitemShut {NoStop}%
\bibitem [{\citenamefont {B{\'e}car}\ \emph {et~al.}(2025)\citenamefont {B{\'e}car}, \citenamefont {Gonz{\'a}lez}, \citenamefont {Papantonopoulos},\ and\ \citenamefont {V{\'a}squez}}]{Becar:2025niq}%
  \BibitemOpen
  \bibfield  {author} {\bibinfo {author} {\bibfnamefont {R.}~\bibnamefont {B{\'e}car}}, \bibinfo {author} {\bibfnamefont {P.~A.}\ \bibnamefont {Gonz{\'a}lez}}, \bibinfo {author} {\bibfnamefont {E.}~\bibnamefont {Papantonopoulos}}, \ and\ \bibinfo {author} {\bibfnamefont {Y.}~\bibnamefont {V{\'a}squez}},\ }\href {\doibase 10.1103/y33s-2hd1} {\bibfield  {journal} {\bibinfo  {journal} {Phys. Rev. D}\ }\textbf {\bibinfo {volume} {111}},\ \bibinfo {pages} {124013} (\bibinfo {year} {2025})},\ \Eprint {http://arxiv.org/abs/2505.17161} {arXiv:2505.17161 [gr-qc]} \BibitemShut {NoStop}%
\bibitem [{\citenamefont {Chen}\ \emph {et~al.}(2025)\citenamefont {Chen}, \citenamefont {Jing}, \citenamefont {Cao},\ and\ \citenamefont {Wang}}]{Chen:2025sbz}%
  \BibitemOpen
  \bibfield  {author} {\bibinfo {author} {\bibfnamefont {C.}~\bibnamefont {Chen}}, \bibinfo {author} {\bibfnamefont {J.}~\bibnamefont {Jing}}, \bibinfo {author} {\bibfnamefont {Z.}~\bibnamefont {Cao}}, \ and\ \bibinfo {author} {\bibfnamefont {M.}~\bibnamefont {Wang}},\ }\href@noop {} {\  (\bibinfo {year} {2025})},\ \Eprint {http://arxiv.org/abs/2506.14635} {arXiv:2506.14635 [gr-qc]} \BibitemShut {NoStop}%
\bibitem [{\citenamefont {Zhang}\ \emph {et~al.}(2019)\citenamefont {Zhang}, \citenamefont {Jiang},\ and\ \citenamefont {Zhong}}]{Zhang:2018jgj}%
  \BibitemOpen
  \bibfield  {author} {\bibinfo {author} {\bibfnamefont {M.}~\bibnamefont {Zhang}}, \bibinfo {author} {\bibfnamefont {J.}~\bibnamefont {Jiang}}, \ and\ \bibinfo {author} {\bibfnamefont {Z.}~\bibnamefont {Zhong}},\ }\href {\doibase 10.1016/j.physletb.2018.10.072} {\bibfield  {journal} {\bibinfo  {journal} {Phys. Lett. B}\ }\textbf {\bibinfo {volume} {789}},\ \bibinfo {pages} {13} (\bibinfo {year} {2019})},\ \Eprint {http://arxiv.org/abs/1811.04183} {arXiv:1811.04183 [gr-qc]} \BibitemShut {NoStop}%
\bibitem [{\citenamefont {Arag\'on}\ \emph {et~al.}(2021)\citenamefont {Arag\'on}, \citenamefont {B\'ecar}, \citenamefont {Gonz\'alez},\ and\ \citenamefont {V\'asquez}}]{Aragon:2020teq}%
  \BibitemOpen
  \bibfield  {author} {\bibinfo {author} {\bibfnamefont {A.}~\bibnamefont {Arag\'on}}, \bibinfo {author} {\bibfnamefont {R.}~\bibnamefont {B\'ecar}}, \bibinfo {author} {\bibfnamefont {P.~A.}\ \bibnamefont {Gonz\'alez}}, \ and\ \bibinfo {author} {\bibfnamefont {Y.}~\bibnamefont {V\'asquez}},\ }\href {\doibase 10.1103/PhysRevD.103.064006} {\bibfield  {journal} {\bibinfo  {journal} {Phys. Rev. D}\ }\textbf {\bibinfo {volume} {103}},\ \bibinfo {pages} {064006} (\bibinfo {year} {2021})},\ \Eprint {http://arxiv.org/abs/2009.09436} {arXiv:2009.09436 [gr-qc]} \BibitemShut {NoStop}%
\bibitem [{\citenamefont {Jing}(2005)}]{Jing:2004zb}%
  \BibitemOpen
  \bibfield  {author} {\bibinfo {author} {\bibfnamefont {J.}~\bibnamefont {Jing}},\ }\href {\doibase 10.1103/PhysRevD.72.027501} {\bibfield  {journal} {\bibinfo  {journal} {Phys. Rev. D}\ }\textbf {\bibinfo {volume} {72}},\ \bibinfo {pages} {027501} (\bibinfo {year} {2005})},\ \Eprint {http://arxiv.org/abs/gr-qc/0408090} {arXiv:gr-qc/0408090} \BibitemShut {NoStop}%
\bibitem [{\citenamefont {Koyama}\ and\ \citenamefont {Tomimatsu}(2002)}]{Koyama:2001qw}%
  \BibitemOpen
  \bibfield  {author} {\bibinfo {author} {\bibfnamefont {H.}~\bibnamefont {Koyama}}\ and\ \bibinfo {author} {\bibfnamefont {A.}~\bibnamefont {Tomimatsu}},\ }\href {\doibase 10.1103/PhysRevD.65.084031} {\bibfield  {journal} {\bibinfo  {journal} {Phys. Rev. D}\ }\textbf {\bibinfo {volume} {65}},\ \bibinfo {pages} {084031} (\bibinfo {year} {2002})},\ \Eprint {http://arxiv.org/abs/gr-qc/0112075} {arXiv:gr-qc/0112075} \BibitemShut {NoStop}%
\bibitem [{\citenamefont {Moderski}\ and\ \citenamefont {Rogatko}(2001)}]{Moderski:2001tk}%
  \BibitemOpen
  \bibfield  {author} {\bibinfo {author} {\bibfnamefont {R.}~\bibnamefont {Moderski}}\ and\ \bibinfo {author} {\bibfnamefont {M.}~\bibnamefont {Rogatko}},\ }\href {\doibase 10.1103/PhysRevD.64.044024} {\bibfield  {journal} {\bibinfo  {journal} {Phys. Rev. D}\ }\textbf {\bibinfo {volume} {64}},\ \bibinfo {pages} {044024} (\bibinfo {year} {2001})},\ \Eprint {http://arxiv.org/abs/gr-qc/0105056} {arXiv:gr-qc/0105056} \BibitemShut {NoStop}%
\bibitem [{\citenamefont {Rogatko}\ and\ \citenamefont {Szyplowska}(2007)}]{Rogatko:2007zz}%
  \BibitemOpen
  \bibfield  {author} {\bibinfo {author} {\bibfnamefont {M.}~\bibnamefont {Rogatko}}\ and\ \bibinfo {author} {\bibfnamefont {A.}~\bibnamefont {Szyplowska}},\ }\href {\doibase 10.1103/PhysRevD.76.044010} {\bibfield  {journal} {\bibinfo  {journal} {Phys. Rev. D}\ }\textbf {\bibinfo {volume} {76}},\ \bibinfo {pages} {044010} (\bibinfo {year} {2007})}\BibitemShut {NoStop}%
\bibitem [{\citenamefont {Koyama}\ and\ \citenamefont {Tomimatsu}(2001{\natexlab{a}})}]{Koyama:2001ee}%
  \BibitemOpen
  \bibfield  {author} {\bibinfo {author} {\bibfnamefont {H.}~\bibnamefont {Koyama}}\ and\ \bibinfo {author} {\bibfnamefont {A.}~\bibnamefont {Tomimatsu}},\ }\href {\doibase 10.1103/PhysRevD.64.044014} {\bibfield  {journal} {\bibinfo  {journal} {Phys. Rev. D}\ }\textbf {\bibinfo {volume} {64}},\ \bibinfo {pages} {044014} (\bibinfo {year} {2001}{\natexlab{a}})},\ \Eprint {http://arxiv.org/abs/gr-qc/0103086} {arXiv:gr-qc/0103086} \BibitemShut {NoStop}%
\bibitem [{\citenamefont {Koyama}\ and\ \citenamefont {Tomimatsu}(2001{\natexlab{b}})}]{Koyama:2000hj}%
  \BibitemOpen
  \bibfield  {author} {\bibinfo {author} {\bibfnamefont {H.}~\bibnamefont {Koyama}}\ and\ \bibinfo {author} {\bibfnamefont {A.}~\bibnamefont {Tomimatsu}},\ }\href {\doibase 10.1103/PhysRevD.63.064032} {\bibfield  {journal} {\bibinfo  {journal} {Phys. Rev. D}\ }\textbf {\bibinfo {volume} {63}},\ \bibinfo {pages} {064032} (\bibinfo {year} {2001}{\natexlab{b}})},\ \Eprint {http://arxiv.org/abs/gr-qc/0012022} {arXiv:gr-qc/0012022} \BibitemShut {NoStop}%
\bibitem [{\citenamefont {Gibbons}\ \emph {et~al.}(2008)\citenamefont {Gibbons}, \citenamefont {Rogatko},\ and\ \citenamefont {Szyplowska}}]{Gibbons:2008gg}%
  \BibitemOpen
  \bibfield  {author} {\bibinfo {author} {\bibfnamefont {G.~W.}\ \bibnamefont {Gibbons}}, \bibinfo {author} {\bibfnamefont {M.}~\bibnamefont {Rogatko}}, \ and\ \bibinfo {author} {\bibfnamefont {A.}~\bibnamefont {Szyplowska}},\ }\href {\doibase 10.1103/PhysRevD.77.064024} {\bibfield  {journal} {\bibinfo  {journal} {Phys. Rev. D}\ }\textbf {\bibinfo {volume} {77}},\ \bibinfo {pages} {064024} (\bibinfo {year} {2008})},\ \Eprint {http://arxiv.org/abs/0802.3259} {arXiv:0802.3259 [hep-th]} \BibitemShut {NoStop}%
\bibitem [{\citenamefont {Gibbons}\ and\ \citenamefont {Rogatko}(2008)}]{Gibbons:2008rs}%
  \BibitemOpen
  \bibfield  {author} {\bibinfo {author} {\bibfnamefont {G.~W.}\ \bibnamefont {Gibbons}}\ and\ \bibinfo {author} {\bibfnamefont {M.}~\bibnamefont {Rogatko}},\ }\href {\doibase 10.1103/PhysRevD.77.044034} {\bibfield  {journal} {\bibinfo  {journal} {Phys. Rev. D}\ }\textbf {\bibinfo {volume} {77}},\ \bibinfo {pages} {044034} (\bibinfo {year} {2008})},\ \Eprint {http://arxiv.org/abs/0801.3130} {arXiv:0801.3130 [hep-th]} \BibitemShut {NoStop}%
\bibitem [{\citenamefont {Konoplya}(2006)}]{Konoplya:2005hr}%
  \BibitemOpen
  \bibfield  {author} {\bibinfo {author} {\bibfnamefont {R.~A.}\ \bibnamefont {Konoplya}},\ }\href {\doibase 10.1103/PhysRevD.73.024009} {\bibfield  {journal} {\bibinfo  {journal} {Phys. Rev. D}\ }\textbf {\bibinfo {volume} {73}},\ \bibinfo {pages} {024009} (\bibinfo {year} {2006})},\ \Eprint {http://arxiv.org/abs/gr-qc/0509026} {arXiv:gr-qc/0509026} \BibitemShut {NoStop}%
\bibitem [{\citenamefont {Fernandes}\ \emph {et~al.}(2022)\citenamefont {Fernandes}, \citenamefont {Hilditch}, \citenamefont {Lemos},\ and\ \citenamefont {Cardoso}}]{Fernandes:2021qvr}%
  \BibitemOpen
  \bibfield  {author} {\bibinfo {author} {\bibfnamefont {T.~V.}\ \bibnamefont {Fernandes}}, \bibinfo {author} {\bibfnamefont {D.}~\bibnamefont {Hilditch}}, \bibinfo {author} {\bibfnamefont {J.~P.~S.}\ \bibnamefont {Lemos}}, \ and\ \bibinfo {author} {\bibfnamefont {V.}~\bibnamefont {Cardoso}},\ }\href {\doibase 10.1103/PhysRevD.105.044017} {\bibfield  {journal} {\bibinfo  {journal} {Phys. Rev. D}\ }\textbf {\bibinfo {volume} {105}},\ \bibinfo {pages} {044017} (\bibinfo {year} {2022})},\ \Eprint {http://arxiv.org/abs/2112.03282} {arXiv:2112.03282 [gr-qc]} \BibitemShut {NoStop}%
\bibitem [{\citenamefont {Konoplya}\ and\ \citenamefont {Zhidenko}(2018)}]{Konoplya:2017tvu}%
  \BibitemOpen
  \bibfield  {author} {\bibinfo {author} {\bibfnamefont {R.~A.}\ \bibnamefont {Konoplya}}\ and\ \bibinfo {author} {\bibfnamefont {A.}~\bibnamefont {Zhidenko}},\ }\href {\doibase 10.1103/PhysRevD.97.084034} {\bibfield  {journal} {\bibinfo  {journal} {Phys. Rev. D}\ }\textbf {\bibinfo {volume} {97}},\ \bibinfo {pages} {084034} (\bibinfo {year} {2018})},\ \Eprint {http://arxiv.org/abs/1712.06667} {arXiv:1712.06667 [gr-qc]} \BibitemShut {NoStop}%
\bibitem [{\citenamefont {Percival}\ and\ \citenamefont {Dolan}(2020)}]{Percival:2020skc}%
  \BibitemOpen
  \bibfield  {author} {\bibinfo {author} {\bibfnamefont {J.}~\bibnamefont {Percival}}\ and\ \bibinfo {author} {\bibfnamefont {S.~R.}\ \bibnamefont {Dolan}},\ }\href {\doibase 10.1103/PhysRevD.102.104055} {\bibfield  {journal} {\bibinfo  {journal} {Phys. Rev. D}\ }\textbf {\bibinfo {volume} {102}},\ \bibinfo {pages} {104055} (\bibinfo {year} {2020})},\ \Eprint {http://arxiv.org/abs/2008.10621} {arXiv:2008.10621 [gr-qc]} \BibitemShut {NoStop}%
\bibitem [{\citenamefont {Zhidenko}(2006)}]{Zhidenko:2006rs}%
  \BibitemOpen
  \bibfield  {author} {\bibinfo {author} {\bibfnamefont {A.}~\bibnamefont {Zhidenko}},\ }\href {\doibase 10.1103/PhysRevD.74.064017} {\bibfield  {journal} {\bibinfo  {journal} {Phys. Rev. D}\ }\textbf {\bibinfo {volume} {74}},\ \bibinfo {pages} {064017} (\bibinfo {year} {2006})},\ \Eprint {http://arxiv.org/abs/gr-qc/0607133} {arXiv:gr-qc/0607133} \BibitemShut {NoStop}%
\bibitem [{\citenamefont {Zinhailo}(2018)}]{Zinhailo:2018ska}%
  \BibitemOpen
  \bibfield  {author} {\bibinfo {author} {\bibfnamefont {A.~F.}\ \bibnamefont {Zinhailo}},\ }\href {\doibase 10.1140/epjc/s10052-018-6467-8} {\bibfield  {journal} {\bibinfo  {journal} {Eur. Phys. J. C}\ }\textbf {\bibinfo {volume} {78}},\ \bibinfo {pages} {992} (\bibinfo {year} {2018})},\ \Eprint {http://arxiv.org/abs/1809.03913} {arXiv:1809.03913 [gr-qc]} \BibitemShut {NoStop}%
\bibitem [{\citenamefont {Konoplya}\ \emph {et~al.}(2019{\natexlab{a}})\citenamefont {Konoplya}, \citenamefont {Zinhailo},\ and\ \citenamefont {Stuchl\'\i{}k}}]{Konoplya:2019hml}%
  \BibitemOpen
  \bibfield  {author} {\bibinfo {author} {\bibfnamefont {R.~A.}\ \bibnamefont {Konoplya}}, \bibinfo {author} {\bibfnamefont {A.~F.}\ \bibnamefont {Zinhailo}}, \ and\ \bibinfo {author} {\bibfnamefont {Z.}~\bibnamefont {Stuchl\'\i{}k}},\ }\href {\doibase 10.1103/PhysRevD.99.124042} {\bibfield  {journal} {\bibinfo  {journal} {Phys. Rev. D}\ }\textbf {\bibinfo {volume} {99}},\ \bibinfo {pages} {124042} (\bibinfo {year} {2019}{\natexlab{a}})},\ \Eprint {http://arxiv.org/abs/1903.03483} {arXiv:1903.03483 [gr-qc]} \BibitemShut {NoStop}%
\bibitem [{\citenamefont {Bolokhov}(2024{\natexlab{a}})}]{Bolokhov:2023bwm}%
  \BibitemOpen
  \bibfield  {author} {\bibinfo {author} {\bibfnamefont {S.~V.}\ \bibnamefont {Bolokhov}},\ }\href {\doibase 10.1103/PhysRevD.110.024010} {\bibfield  {journal} {\bibinfo  {journal} {Phys. Rev. D}\ }\textbf {\bibinfo {volume} {110}},\ \bibinfo {pages} {024010} (\bibinfo {year} {2024}{\natexlab{a}})},\ \Eprint {http://arxiv.org/abs/2311.05503} {arXiv:2311.05503 [gr-qc]} \BibitemShut {NoStop}%
\bibitem [{\citenamefont {Churilova}\ \emph {et~al.}(2020)\citenamefont {Churilova}, \citenamefont {Konoplya},\ and\ \citenamefont {Zhidenko}}]{Churilova:2019qph}%
  \BibitemOpen
  \bibfield  {author} {\bibinfo {author} {\bibfnamefont {M.~S.}\ \bibnamefont {Churilova}}, \bibinfo {author} {\bibfnamefont {R.~A.}\ \bibnamefont {Konoplya}}, \ and\ \bibinfo {author} {\bibfnamefont {A.}~\bibnamefont {Zhidenko}},\ }\href {\doibase 10.1016/j.physletb.2020.135207} {\bibfield  {journal} {\bibinfo  {journal} {Phys. Lett. B}\ }\textbf {\bibinfo {volume} {802}},\ \bibinfo {pages} {135207} (\bibinfo {year} {2020})},\ \Eprint {http://arxiv.org/abs/1911.05246} {arXiv:1911.05246 [gr-qc]} \BibitemShut {NoStop}%
\bibitem [{\citenamefont {Seahra}\ \emph {et~al.}(2005)\citenamefont {Seahra}, \citenamefont {Clarkson},\ and\ \citenamefont {Maartens}}]{Seahra:2004fg}%
  \BibitemOpen
  \bibfield  {author} {\bibinfo {author} {\bibfnamefont {S.~S.}\ \bibnamefont {Seahra}}, \bibinfo {author} {\bibfnamefont {C.}~\bibnamefont {Clarkson}}, \ and\ \bibinfo {author} {\bibfnamefont {R.}~\bibnamefont {Maartens}},\ }\href {\doibase 10.1103/PhysRevLett.94.121302} {\bibfield  {journal} {\bibinfo  {journal} {Phys. Rev. Lett.}\ }\textbf {\bibinfo {volume} {94}},\ \bibinfo {pages} {121302} (\bibinfo {year} {2005})},\ \Eprint {http://arxiv.org/abs/gr-qc/0408032} {arXiv:gr-qc/0408032} \BibitemShut {NoStop}%
\bibitem [{\citenamefont {Konoplya}\ and\ \citenamefont {Fontana}(2008)}]{Konoplya:2007yy}%
  \BibitemOpen
  \bibfield  {author} {\bibinfo {author} {\bibfnamefont {R.~A.}\ \bibnamefont {Konoplya}}\ and\ \bibinfo {author} {\bibfnamefont {R.~D.~B.}\ \bibnamefont {Fontana}},\ }\href {\doibase 10.1016/j.physletb.2007.10.065} {\bibfield  {journal} {\bibinfo  {journal} {Phys. Lett. B}\ }\textbf {\bibinfo {volume} {659}},\ \bibinfo {pages} {375} (\bibinfo {year} {2008})},\ \Eprint {http://arxiv.org/abs/0707.1156} {arXiv:0707.1156 [hep-th]} \BibitemShut {NoStop}%
\bibitem [{\citenamefont {Konoplya}(2008)}]{Konoplya:2008hj}%
  \BibitemOpen
  \bibfield  {author} {\bibinfo {author} {\bibfnamefont {R.~A.}\ \bibnamefont {Konoplya}},\ }\href {\doibase 10.1016/j.physletb.2008.11.059} {\bibfield  {journal} {\bibinfo  {journal} {Phys. Lett. B}\ }\textbf {\bibinfo {volume} {666}},\ \bibinfo {pages} {283} (\bibinfo {year} {2008})},\ \Eprint {http://arxiv.org/abs/0801.0846} {arXiv:0801.0846 [hep-th]} \BibitemShut {NoStop}%
\bibitem [{\citenamefont {Wu}\ and\ \citenamefont {Xu}(2015)}]{Wu:2015fwa}%
  \BibitemOpen
  \bibfield  {author} {\bibinfo {author} {\bibfnamefont {C.}~\bibnamefont {Wu}}\ and\ \bibinfo {author} {\bibfnamefont {R.}~\bibnamefont {Xu}},\ }\href {\doibase 10.1140/epjc/s10052-015-3632-1} {\bibfield  {journal} {\bibinfo  {journal} {Eur. Phys. J. C}\ }\textbf {\bibinfo {volume} {75}},\ \bibinfo {pages} {391} (\bibinfo {year} {2015})},\ \Eprint {http://arxiv.org/abs/1507.04911} {arXiv:1507.04911 [gr-qc]} \BibitemShut {NoStop}%
\bibitem [{\citenamefont {Kokkotas}\ \emph {et~al.}(2011)\citenamefont {Kokkotas}, \citenamefont {Konoplya},\ and\ \citenamefont {Zhidenko}}]{Kokkotas:2010zd}%
  \BibitemOpen
  \bibfield  {author} {\bibinfo {author} {\bibfnamefont {K.~D.}\ \bibnamefont {Kokkotas}}, \bibinfo {author} {\bibfnamefont {R.~A.}\ \bibnamefont {Konoplya}}, \ and\ \bibinfo {author} {\bibfnamefont {A.}~\bibnamefont {Zhidenko}},\ }\href {\doibase 10.1103/PhysRevD.83.024031} {\bibfield  {journal} {\bibinfo  {journal} {Phys. Rev. D}\ }\textbf {\bibinfo {volume} {83}},\ \bibinfo {pages} {024031} (\bibinfo {year} {2011})},\ \Eprint {http://arxiv.org/abs/1011.1843} {arXiv:1011.1843 [gr-qc]} \BibitemShut {NoStop}%
\bibitem [{\citenamefont {Konoplya}\ and\ \citenamefont {Zhidenko}(2024)}]{Konoplya:2023fmh}%
  \BibitemOpen
  \bibfield  {author} {\bibinfo {author} {\bibfnamefont {R.~A.}\ \bibnamefont {Konoplya}}\ and\ \bibinfo {author} {\bibfnamefont {A.}~\bibnamefont {Zhidenko}},\ }\href {\doibase 10.1016/j.physletb.2024.138685} {\bibfield  {journal} {\bibinfo  {journal} {Phys. Lett. B}\ }\textbf {\bibinfo {volume} {853}},\ \bibinfo {pages} {138685} (\bibinfo {year} {2024})},\ \Eprint {http://arxiv.org/abs/2307.01110} {arXiv:2307.01110 [gr-qc]} \BibitemShut {NoStop}%
\bibitem [{\citenamefont {Afzal}\ \emph {et~al.}(2023)\citenamefont {Afzal} \emph {et~al.}}]{NANOGrav:2023hvm}%
  \BibitemOpen
  \bibfield  {author} {\bibinfo {author} {\bibfnamefont {A.}~\bibnamefont {Afzal}} \emph {et~al.} (\bibinfo {collaboration} {NANOGrav}),\ }\href {\doibase 10.3847/2041-8213/acdc91} {\bibfield  {journal} {\bibinfo  {journal} {Astrophys. J. Lett.}\ }\textbf {\bibinfo {volume} {951}},\ \bibinfo {pages} {L11} (\bibinfo {year} {2023})},\ \Eprint {http://arxiv.org/abs/2306.16219} {arXiv:2306.16219 [astro-ph.HE]} \BibitemShut {NoStop}%
\bibitem [{\citenamefont {Zinhailo}(2024{\natexlab{a}})}]{Zinhailo:2024jzt}%
  \BibitemOpen
  \bibfield  {author} {\bibinfo {author} {\bibfnamefont {A.~F.}\ \bibnamefont {Zinhailo}},\ }\href {\doibase 10.1016/j.physletb.2024.138682} {\bibfield  {journal} {\bibinfo  {journal} {Phys. Lett. B}\ }\textbf {\bibinfo {volume} {853}},\ \bibinfo {pages} {138682} (\bibinfo {year} {2024}{\natexlab{a}})},\ \Eprint {http://arxiv.org/abs/2403.06867} {arXiv:2403.06867 [gr-qc]} \BibitemShut {NoStop}%
\bibitem [{\citenamefont {Cardoso}\ \emph {et~al.}(2016{\natexlab{a}})\citenamefont {Cardoso}, \citenamefont {Franzin},\ and\ \citenamefont {Pani}}]{Cardoso:2016rao}%
  \BibitemOpen
  \bibfield  {author} {\bibinfo {author} {\bibfnamefont {V.}~\bibnamefont {Cardoso}}, \bibinfo {author} {\bibfnamefont {E.}~\bibnamefont {Franzin}}, \ and\ \bibinfo {author} {\bibfnamefont {P.}~\bibnamefont {Pani}},\ }\href {\doibase 10.1103/PhysRevLett.116.171101} {\bibfield  {journal} {\bibinfo  {journal} {Phys. Rev. Lett.}\ }\textbf {\bibinfo {volume} {116}},\ \bibinfo {pages} {171101} (\bibinfo {year} {2016}{\natexlab{a}})},\ \bibinfo {note} {[Erratum: Phys.Rev.Lett. 117, 089902 (2016)]},\ \Eprint {http://arxiv.org/abs/1602.07309} {arXiv:1602.07309 [gr-qc]} \BibitemShut {NoStop}%
\bibitem [{\citenamefont {Cardoso}\ \emph {et~al.}(2016{\natexlab{b}})\citenamefont {Cardoso}, \citenamefont {Hopper}, \citenamefont {Macedo}, \citenamefont {Palenzuela},\ and\ \citenamefont {Pani}}]{Cardoso:2016oxy}%
  \BibitemOpen
  \bibfield  {author} {\bibinfo {author} {\bibfnamefont {V.}~\bibnamefont {Cardoso}}, \bibinfo {author} {\bibfnamefont {S.}~\bibnamefont {Hopper}}, \bibinfo {author} {\bibfnamefont {C.~F.~B.}\ \bibnamefont {Macedo}}, \bibinfo {author} {\bibfnamefont {C.}~\bibnamefont {Palenzuela}}, \ and\ \bibinfo {author} {\bibfnamefont {P.}~\bibnamefont {Pani}},\ }\href {\doibase 10.1103/PhysRevD.94.084031} {\bibfield  {journal} {\bibinfo  {journal} {Phys. Rev. D}\ }\textbf {\bibinfo {volume} {94}},\ \bibinfo {pages} {084031} (\bibinfo {year} {2016}{\natexlab{b}})},\ \Eprint {http://arxiv.org/abs/1608.08637} {arXiv:1608.08637 [gr-qc]} \BibitemShut {NoStop}%
\bibitem [{\citenamefont {Abedi}\ \emph {et~al.}(2017)\citenamefont {Abedi}, \citenamefont {Dykaar},\ and\ \citenamefont {Afshordi}}]{Abedi:2016hgu}%
  \BibitemOpen
  \bibfield  {author} {\bibinfo {author} {\bibfnamefont {J.}~\bibnamefont {Abedi}}, \bibinfo {author} {\bibfnamefont {H.}~\bibnamefont {Dykaar}}, \ and\ \bibinfo {author} {\bibfnamefont {N.}~\bibnamefont {Afshordi}},\ }\href {\doibase 10.1103/PhysRevD.96.082004} {\bibfield  {journal} {\bibinfo  {journal} {Phys. Rev. D}\ }\textbf {\bibinfo {volume} {96}},\ \bibinfo {pages} {082004} (\bibinfo {year} {2017})},\ \Eprint {http://arxiv.org/abs/1612.00266} {arXiv:1612.00266 [gr-qc]} \BibitemShut {NoStop}%
\bibitem [{\citenamefont {Mark}\ \emph {et~al.}(2017)\citenamefont {Mark}, \citenamefont {Zimmerman}, \citenamefont {Du},\ and\ \citenamefont {Chen}}]{Mark:2017dnq}%
  \BibitemOpen
  \bibfield  {author} {\bibinfo {author} {\bibfnamefont {Z.}~\bibnamefont {Mark}}, \bibinfo {author} {\bibfnamefont {A.}~\bibnamefont {Zimmerman}}, \bibinfo {author} {\bibfnamefont {S.~M.}\ \bibnamefont {Du}}, \ and\ \bibinfo {author} {\bibfnamefont {Y.}~\bibnamefont {Chen}},\ }\href {\doibase 10.1103/PhysRevD.96.084002} {\bibfield  {journal} {\bibinfo  {journal} {Phys. Rev. D}\ }\textbf {\bibinfo {volume} {96}},\ \bibinfo {pages} {084002} (\bibinfo {year} {2017})},\ \Eprint {http://arxiv.org/abs/1706.06155} {arXiv:1706.06155 [gr-qc]} \BibitemShut {NoStop}%
\bibitem [{\citenamefont {Bueno}\ \emph {et~al.}(2018)\citenamefont {Bueno}, \citenamefont {Cano}, \citenamefont {Goelen}, \citenamefont {Hertog},\ and\ \citenamefont {Vercnocke}}]{Bueno:2017hyj}%
  \BibitemOpen
  \bibfield  {author} {\bibinfo {author} {\bibfnamefont {P.}~\bibnamefont {Bueno}}, \bibinfo {author} {\bibfnamefont {P.~A.}\ \bibnamefont {Cano}}, \bibinfo {author} {\bibfnamefont {F.}~\bibnamefont {Goelen}}, \bibinfo {author} {\bibfnamefont {T.}~\bibnamefont {Hertog}}, \ and\ \bibinfo {author} {\bibfnamefont {B.}~\bibnamefont {Vercnocke}},\ }\href {\doibase 10.1103/PhysRevD.97.024040} {\bibfield  {journal} {\bibinfo  {journal} {Phys. Rev. D}\ }\textbf {\bibinfo {volume} {97}},\ \bibinfo {pages} {024040} (\bibinfo {year} {2018})},\ \Eprint {http://arxiv.org/abs/1711.00391} {arXiv:1711.00391 [gr-qc]} \BibitemShut {NoStop}%
\bibitem [{\citenamefont {Konoplya}\ \emph {et~al.}(2019{\natexlab{b}})\citenamefont {Konoplya}, \citenamefont {Stuchlík},\ and\ \citenamefont {Zhidenko}}]{Konoplya:2018yrp}%
  \BibitemOpen
  \bibfield  {author} {\bibinfo {author} {\bibfnamefont {R.~A.}\ \bibnamefont {Konoplya}}, \bibinfo {author} {\bibfnamefont {Z.}~\bibnamefont {Stuchlík}}, \ and\ \bibinfo {author} {\bibfnamefont {A.}~\bibnamefont {Zhidenko}},\ }\href {\doibase 10.1103/PhysRevD.99.024007} {\bibfield  {journal} {\bibinfo  {journal} {Phys. Rev. D}\ }\textbf {\bibinfo {volume} {99}},\ \bibinfo {pages} {024007} (\bibinfo {year} {2019}{\natexlab{b}})},\ \Eprint {http://arxiv.org/abs/1810.01295} {arXiv:1810.01295 [gr-qc]} \BibitemShut {NoStop}%
\bibitem [{\citenamefont {Cardoso}\ \emph {et~al.}(2019)\citenamefont {Cardoso}, \citenamefont {Foit},\ and\ \citenamefont {Kleban}}]{Cardoso:2019apo}%
  \BibitemOpen
  \bibfield  {author} {\bibinfo {author} {\bibfnamefont {V.}~\bibnamefont {Cardoso}}, \bibinfo {author} {\bibfnamefont {V.~F.}\ \bibnamefont {Foit}}, \ and\ \bibinfo {author} {\bibfnamefont {M.}~\bibnamefont {Kleban}},\ }\href {\doibase 10.1088/1475-7516/2019/08/006} {\bibfield  {journal} {\bibinfo  {journal} {JCAP}\ }\textbf {\bibinfo {volume} {08}},\ \bibinfo {pages} {006} (\bibinfo {year} {2019})},\ \Eprint {http://arxiv.org/abs/1902.10164} {arXiv:1902.10164 [hep-th]} \BibitemShut {NoStop}%
\bibitem [{\citenamefont {Bronnikov}\ and\ \citenamefont {Konoplya}(2020)}]{Bronnikov:2019sbx}%
  \BibitemOpen
  \bibfield  {author} {\bibinfo {author} {\bibfnamefont {K.~A.}\ \bibnamefont {Bronnikov}}\ and\ \bibinfo {author} {\bibfnamefont {R.~A.}\ \bibnamefont {Konoplya}},\ }\href {\doibase 10.1103/PhysRevD.101.064004} {\bibfield  {journal} {\bibinfo  {journal} {Phys. Rev. D}\ }\textbf {\bibinfo {volume} {101}},\ \bibinfo {pages} {064004} (\bibinfo {year} {2020})},\ \Eprint {http://arxiv.org/abs/1912.05315} {arXiv:1912.05315 [gr-qc]} \BibitemShut {NoStop}%
\bibitem [{\citenamefont {Dong}\ and\ \citenamefont {Stojkovic}(2021)}]{Dong:2020odp}%
  \BibitemOpen
  \bibfield  {author} {\bibinfo {author} {\bibfnamefont {R.}~\bibnamefont {Dong}}\ and\ \bibinfo {author} {\bibfnamefont {D.}~\bibnamefont {Stojkovic}},\ }\href {\doibase 10.1103/PhysRevD.103.024058} {\bibfield  {journal} {\bibinfo  {journal} {Phys. Rev. D}\ }\textbf {\bibinfo {volume} {103}},\ \bibinfo {pages} {024058} (\bibinfo {year} {2021})},\ \Eprint {http://arxiv.org/abs/2011.04032} {arXiv:2011.04032 [gr-qc]} \BibitemShut {NoStop}%
\bibitem [{\citenamefont {Churilova}\ \emph {et~al.}(2021)\citenamefont {Churilova}, \citenamefont {Konoplya}, \citenamefont {Stuchlík},\ and\ \citenamefont {Zhidenko}}]{Churilova:2021tgn}%
  \BibitemOpen
  \bibfield  {author} {\bibinfo {author} {\bibfnamefont {M.~S.}\ \bibnamefont {Churilova}}, \bibinfo {author} {\bibfnamefont {R.~A.}\ \bibnamefont {Konoplya}}, \bibinfo {author} {\bibfnamefont {Z.}~\bibnamefont {Stuchlík}}, \ and\ \bibinfo {author} {\bibfnamefont {A.}~\bibnamefont {Zhidenko}},\ }\href {\doibase 10.1088/1475-7516/2021/10/010} {\bibfield  {journal} {\bibinfo  {journal} {JCAP}\ }\textbf {\bibinfo {volume} {10}},\ \bibinfo {pages} {010} (\bibinfo {year} {2021})},\ \Eprint {http://arxiv.org/abs/2107.05977} {arXiv:2107.05977 [gr-qc]} \BibitemShut {NoStop}%
\bibitem [{\citenamefont {Charmousis}\ \emph {et~al.}(2025)\citenamefont {Charmousis}, \citenamefont {Fernandes},\ and\ \citenamefont {Hassaine}}]{Charmousis:2025jpx}%
  \BibitemOpen
  \bibfield  {author} {\bibinfo {author} {\bibfnamefont {C.}~\bibnamefont {Charmousis}}, \bibinfo {author} {\bibfnamefont {P.~G.~S.}\ \bibnamefont {Fernandes}}, \ and\ \bibinfo {author} {\bibfnamefont {M.}~\bibnamefont {Hassaine}},\ }\href {\doibase 10.1103/9f2w-3kly} {\bibfield  {journal} {\bibinfo  {journal} {Phys. Rev. D}\ }\textbf {\bibinfo {volume} {111}},\ \bibinfo {pages} {124008} (\bibinfo {year} {2025})},\ \Eprint {http://arxiv.org/abs/2504.13084} {arXiv:2504.13084 [gr-qc]} \BibitemShut {NoStop}%
\bibitem [{\citenamefont {Konoplya}\ and\ \citenamefont {Zinhailo}(2020{\natexlab{a}})}]{Konoplya:2020cbv}%
  \BibitemOpen
  \bibfield  {author} {\bibinfo {author} {\bibfnamefont {R.~A.}\ \bibnamefont {Konoplya}}\ and\ \bibinfo {author} {\bibfnamefont {A.~F.}\ \bibnamefont {Zinhailo}},\ }\href {\doibase 10.1016/j.physletb.2020.135793} {\bibfield  {journal} {\bibinfo  {journal} {Phys. Lett. B}\ }\textbf {\bibinfo {volume} {810}},\ \bibinfo {pages} {135793} (\bibinfo {year} {2020}{\natexlab{a}})},\ \Eprint {http://arxiv.org/abs/2004.02248} {arXiv:2004.02248 [gr-qc]} \BibitemShut {NoStop}%
\bibitem [{\citenamefont {Cuyubamba}(2021)}]{Cuyubamba:2020moe}%
  \BibitemOpen
  \bibfield  {author} {\bibinfo {author} {\bibfnamefont {M.~A.}\ \bibnamefont {Cuyubamba}},\ }\href {\doibase 10.1016/j.dark.2021.100789} {\bibfield  {journal} {\bibinfo  {journal} {Phys. Dark Univ.}\ }\textbf {\bibinfo {volume} {31}},\ \bibinfo {pages} {100789} (\bibinfo {year} {2021})},\ \Eprint {http://arxiv.org/abs/2004.09025} {arXiv:2004.09025 [gr-qc]} \BibitemShut {NoStop}%
\bibitem [{\citenamefont {Konoplya}\ and\ \citenamefont {Zinhailo}(2020{\natexlab{b}})}]{Konoplya:2020bxa}%
  \BibitemOpen
  \bibfield  {author} {\bibinfo {author} {\bibfnamefont {R.~A.}\ \bibnamefont {Konoplya}}\ and\ \bibinfo {author} {\bibfnamefont {A.~F.}\ \bibnamefont {Zinhailo}},\ }\href {\doibase 10.1140/epjc/s10052-020-08639-8} {\bibfield  {journal} {\bibinfo  {journal} {Eur. Phys. J. C}\ }\textbf {\bibinfo {volume} {80}},\ \bibinfo {pages} {1049} (\bibinfo {year} {2020}{\natexlab{b}})},\ \Eprint {http://arxiv.org/abs/2003.01188} {arXiv:2003.01188 [gr-qc]} \BibitemShut {NoStop}%
\bibitem [{\citenamefont {Lütfüoğlu}(2025{\natexlab{a}})}]{Lutfuoglu:2025ldc}%
  \BibitemOpen
  \bibfield  {author} {\bibinfo {author} {\bibfnamefont {B.~C.}\ \bibnamefont {Lütfüoğlu}},\ }\href@noop {} {\bibfield  {journal} {\bibinfo  {journal} {International Journal of Gravitation and Theoretical Physics}\ }\textbf {\bibinfo {volume} {1}},\ \bibinfo {pages} {4} (\bibinfo {year} {2025}{\natexlab{a}})},\ \Eprint {http://arxiv.org/abs/2507.09246} {arXiv:2507.09246 [gr-qc]} \BibitemShut {NoStop}%
\bibitem [{\citenamefont {Konoplya}\ and\ \citenamefont {Zhidenko}(2025)}]{Konoplya:2025uiq}%
  \BibitemOpen
  \bibfield  {author} {\bibinfo {author} {\bibfnamefont {R.~A.}\ \bibnamefont {Konoplya}}\ and\ \bibinfo {author} {\bibfnamefont {A.}~\bibnamefont {Zhidenko}},\ }\href@noop {} {\  (\bibinfo {year} {2025})},\ \Eprint {http://arxiv.org/abs/2508.13069} {arXiv:2508.13069 [gr-qc]} \BibitemShut {NoStop}%
\bibitem [{\citenamefont {Konoplya}\ \emph {et~al.}(2024)\citenamefont {Konoplya}, \citenamefont {Stuchlík},\ and\ \citenamefont {Zhidenko}}]{Konoplya:2024wds}%
  \BibitemOpen
  \bibfield  {author} {\bibinfo {author} {\bibfnamefont {R.~A.}\ \bibnamefont {Konoplya}}, \bibinfo {author} {\bibfnamefont {Z.}~\bibnamefont {Stuchlík}}, \ and\ \bibinfo {author} {\bibfnamefont {A.}~\bibnamefont {Zhidenko}},\ }\href@noop {} {\  (\bibinfo {year} {2024})},\ \Eprint {http://arxiv.org/abs/2411.09014} {arXiv:2411.09014 [gr-qc]} \BibitemShut {NoStop}%
\bibitem [{\citenamefont {Lu}\ and\ \citenamefont {Pang}(2020)}]{Lu:2020iav}%
  \BibitemOpen
  \bibfield  {author} {\bibinfo {author} {\bibfnamefont {H.}~\bibnamefont {Lu}}\ and\ \bibinfo {author} {\bibfnamefont {Y.}~\bibnamefont {Pang}},\ }\href {\doibase 10.1016/j.physletb.2020.135717} {\bibfield  {journal} {\bibinfo  {journal} {Phys. Lett. B}\ }\textbf {\bibinfo {volume} {809}},\ \bibinfo {pages} {135717} (\bibinfo {year} {2020})},\ \Eprint {http://arxiv.org/abs/2003.11552} {arXiv:2003.11552 [gr-qc]} \BibitemShut {NoStop}%
\bibitem [{\citenamefont {Kobayashi}(2020)}]{Kobayashi:2020wqy}%
  \BibitemOpen
  \bibfield  {author} {\bibinfo {author} {\bibfnamefont {T.}~\bibnamefont {Kobayashi}},\ }\href {\doibase 10.1088/1475-7516/2020/07/013} {\bibfield  {journal} {\bibinfo  {journal} {JCAP}\ }\textbf {\bibinfo {volume} {07}},\ \bibinfo {pages} {013} (\bibinfo {year} {2020})},\ \Eprint {http://arxiv.org/abs/2003.12771} {arXiv:2003.12771 [gr-qc]} \BibitemShut {NoStop}%
\bibitem [{\citenamefont {Fernandes}\ \emph {et~al.}(2020)\citenamefont {Fernandes}, \citenamefont {Carrilho}, \citenamefont {Clifton},\ and\ \citenamefont {Mulryne}}]{Fernandes:2020nbq}%
  \BibitemOpen
  \bibfield  {author} {\bibinfo {author} {\bibfnamefont {P.~G.~S.}\ \bibnamefont {Fernandes}}, \bibinfo {author} {\bibfnamefont {P.}~\bibnamefont {Carrilho}}, \bibinfo {author} {\bibfnamefont {T.}~\bibnamefont {Clifton}}, \ and\ \bibinfo {author} {\bibfnamefont {D.~J.}\ \bibnamefont {Mulryne}},\ }\href {\doibase 10.1103/PhysRevD.102.024025} {\bibfield  {journal} {\bibinfo  {journal} {Phys. Rev. D}\ }\textbf {\bibinfo {volume} {102}},\ \bibinfo {pages} {024025} (\bibinfo {year} {2020})},\ \Eprint {http://arxiv.org/abs/2004.08362} {arXiv:2004.08362 [gr-qc]} \BibitemShut {NoStop}%
\bibitem [{\citenamefont {Gundlach}\ \emph {et~al.}(1994)\citenamefont {Gundlach}, \citenamefont {Price},\ and\ \citenamefont {Pullin}}]{Gundlach:1993tp}%
  \BibitemOpen
  \bibfield  {author} {\bibinfo {author} {\bibfnamefont {C.}~\bibnamefont {Gundlach}}, \bibinfo {author} {\bibfnamefont {R.~H.}\ \bibnamefont {Price}}, \ and\ \bibinfo {author} {\bibfnamefont {J.}~\bibnamefont {Pullin}},\ }\href {\doibase 10.1103/PhysRevD.49.883} {\bibfield  {journal} {\bibinfo  {journal} {Phys. Rev. D}\ }\textbf {\bibinfo {volume} {49}},\ \bibinfo {pages} {883} (\bibinfo {year} {1994})},\ \Eprint {http://arxiv.org/abs/gr-qc/9307009} {arXiv:gr-qc/9307009} \BibitemShut {NoStop}%
\bibitem [{\citenamefont {Zhidenko}(2008)}]{Zhidenko:2008fp}%
  \BibitemOpen
  \bibfield  {author} {\bibinfo {author} {\bibfnamefont {A.}~\bibnamefont {Zhidenko}},\ }\href {\doibase 10.1103/PhysRevD.78.024007} {\bibfield  {journal} {\bibinfo  {journal} {Phys. Rev. D}\ }\textbf {\bibinfo {volume} {78}},\ \bibinfo {pages} {024007} (\bibinfo {year} {2008})},\ \Eprint {http://arxiv.org/abs/0802.2262} {arXiv:0802.2262 [gr-qc]} \BibitemShut {NoStop}%
\bibitem [{\citenamefont {Konoplya}\ \emph {et~al.}(2020)\citenamefont {Konoplya}, \citenamefont {Zinhailo},\ and\ \citenamefont {Stuchlik}}]{Konoplya:2020jgt}%
  \BibitemOpen
  \bibfield  {author} {\bibinfo {author} {\bibfnamefont {R.~A.}\ \bibnamefont {Konoplya}}, \bibinfo {author} {\bibfnamefont {A.~F.}\ \bibnamefont {Zinhailo}}, \ and\ \bibinfo {author} {\bibfnamefont {Z.}~\bibnamefont {Stuchlik}},\ }\href {\doibase 10.1103/PhysRevD.102.044023} {\bibfield  {journal} {\bibinfo  {journal} {Phys. Rev. D}\ }\textbf {\bibinfo {volume} {102}},\ \bibinfo {pages} {044023} (\bibinfo {year} {2020})},\ \Eprint {http://arxiv.org/abs/2006.10462} {arXiv:2006.10462 [gr-qc]} \BibitemShut {NoStop}%
\bibitem [{\citenamefont {Lütfüoğlu}(2025{\natexlab{b}})}]{Lutfuoglu:2025hwh}%
  \BibitemOpen
  \bibfield  {author} {\bibinfo {author} {\bibfnamefont {B.~C.}\ \bibnamefont {Lütfüoğlu}},\ }\href {\doibase 10.1088/1475-7516/2025/06/057} {\bibfield  {journal} {\bibinfo  {journal} {JCAP}\ }\textbf {\bibinfo {volume} {06}},\ \bibinfo {pages} {057} (\bibinfo {year} {2025}{\natexlab{b}})},\ \Eprint {http://arxiv.org/abs/2504.09323} {arXiv:2504.09323 [gr-qc]} \BibitemShut {NoStop}%
\bibitem [{\citenamefont {Lütfüoğlu}(2025{\natexlab{c}})}]{Lutfuoglu:2025hjy}%
  \BibitemOpen
  \bibfield  {author} {\bibinfo {author} {\bibfnamefont {B.~C.}\ \bibnamefont {Lütfüoğlu}},\ }\href {\doibase 10.1140/epjc/s10052-025-14210-0} {\bibfield  {journal} {\bibinfo  {journal} {Eur. Phys. J. C}\ }\textbf {\bibinfo {volume} {85}},\ \bibinfo {pages} {486} (\bibinfo {year} {2025}{\natexlab{c}})},\ \Eprint {http://arxiv.org/abs/2503.16087} {arXiv:2503.16087 [gr-qc]} \BibitemShut {NoStop}%
\bibitem [{\citenamefont {Konoplya}\ and\ \citenamefont {Zhidenko}(2020)}]{Konoplya:2020juj}%
  \BibitemOpen
  \bibfield  {author} {\bibinfo {author} {\bibfnamefont {R.~A.}\ \bibnamefont {Konoplya}}\ and\ \bibinfo {author} {\bibfnamefont {A.}~\bibnamefont {Zhidenko}},\ }\href {\doibase 10.1016/j.dark.2020.100697} {\bibfield  {journal} {\bibinfo  {journal} {Phys. Dark Univ.}\ }\textbf {\bibinfo {volume} {30}},\ \bibinfo {pages} {100697} (\bibinfo {year} {2020})},\ \Eprint {http://arxiv.org/abs/2003.12492} {arXiv:2003.12492 [gr-qc]} \BibitemShut {NoStop}%
\bibitem [{\citenamefont {Skvortsova}(2024{\natexlab{a}})}]{Skvortsova:2024atk}%
  \BibitemOpen
  \bibfield  {author} {\bibinfo {author} {\bibfnamefont {M.}~\bibnamefont {Skvortsova}},\ }\href {\doibase 10.1002/prop.202400132} {\bibfield  {journal} {\bibinfo  {journal} {Fortsch. Phys.}\ }\textbf {\bibinfo {volume} {72}},\ \bibinfo {pages} {2400132} (\bibinfo {year} {2024}{\natexlab{a}})},\ \Eprint {http://arxiv.org/abs/2405.06390} {arXiv:2405.06390 [gr-qc]} \BibitemShut {NoStop}%
\bibitem [{\citenamefont {Skvortsova}(2024{\natexlab{b}})}]{Skvortsova:2023zmj}%
  \BibitemOpen
  \bibfield  {author} {\bibinfo {author} {\bibfnamefont {M.}~\bibnamefont {Skvortsova}},\ }\href {\doibase 10.1002/prop.202400036} {\bibfield  {journal} {\bibinfo  {journal} {Fortsch. Phys.}\ }\textbf {\bibinfo {volume} {72}},\ \bibinfo {pages} {2400036} (\bibinfo {year} {2024}{\natexlab{b}})},\ \Eprint {http://arxiv.org/abs/2311.11650} {arXiv:2311.11650 [gr-qc]} \BibitemShut {NoStop}%
\bibitem [{\citenamefont {Malik}(2025{\natexlab{a}})}]{Malik:2024qsz}%
  \BibitemOpen
  \bibfield  {author} {\bibinfo {author} {\bibfnamefont {Z.}~\bibnamefont {Malik}},\ }\href {\doibase 10.1142/S0217751X2450132X} {\bibfield  {journal} {\bibinfo  {journal} {Int. J. Mod. Phys. A}\ }\textbf {\bibinfo {volume} {40}},\ \bibinfo {pages} {2450132} (\bibinfo {year} {2025}{\natexlab{a}})}\BibitemShut {NoStop}%
\bibitem [{\citenamefont {Schutz}\ and\ \citenamefont {Will}(1985)}]{Schutz:1985km}%
  \BibitemOpen
  \bibfield  {author} {\bibinfo {author} {\bibfnamefont {B.~F.}\ \bibnamefont {Schutz}}\ and\ \bibinfo {author} {\bibfnamefont {C.~M.}\ \bibnamefont {Will}},\ }\href {\doibase 10.1086/184453} {\bibfield  {journal} {\bibinfo  {journal} {Astrophys. J. Lett.}\ }\textbf {\bibinfo {volume} {291}},\ \bibinfo {pages} {L33} (\bibinfo {year} {1985})}\BibitemShut {NoStop}%
\bibitem [{\citenamefont {Iyer}\ and\ \citenamefont {Will}(1987)}]{Iyer:1986np}%
  \BibitemOpen
  \bibfield  {author} {\bibinfo {author} {\bibfnamefont {S.}~\bibnamefont {Iyer}}\ and\ \bibinfo {author} {\bibfnamefont {C.~M.}\ \bibnamefont {Will}},\ }\href {\doibase 10.1103/PhysRevD.35.3621} {\bibfield  {journal} {\bibinfo  {journal} {Phys. Rev. D}\ }\textbf {\bibinfo {volume} {35}},\ \bibinfo {pages} {3621} (\bibinfo {year} {1987})}\BibitemShut {NoStop}%
\bibitem [{\citenamefont {Konoplya}(2003)}]{Konoplya:2003ii}%
  \BibitemOpen
  \bibfield  {author} {\bibinfo {author} {\bibfnamefont {R.~A.}\ \bibnamefont {Konoplya}},\ }\href {\doibase 10.1103/PhysRevD.68.024018} {\bibfield  {journal} {\bibinfo  {journal} {Phys. Rev. D}\ }\textbf {\bibinfo {volume} {68}},\ \bibinfo {pages} {024018} (\bibinfo {year} {2003})},\ \Eprint {http://arxiv.org/abs/gr-qc/0303052} {arXiv:gr-qc/0303052} \BibitemShut {NoStop}%
\bibitem [{\citenamefont {Matyjasek}\ and\ \citenamefont {Opala}(2017)}]{Matyjasek:2017psv}%
  \BibitemOpen
  \bibfield  {author} {\bibinfo {author} {\bibfnamefont {J.}~\bibnamefont {Matyjasek}}\ and\ \bibinfo {author} {\bibfnamefont {M.}~\bibnamefont {Opala}},\ }\href {\doibase 10.1103/PhysRevD.96.024011} {\bibfield  {journal} {\bibinfo  {journal} {Phys. Rev. D}\ }\textbf {\bibinfo {volume} {96}},\ \bibinfo {pages} {024011} (\bibinfo {year} {2017})},\ \Eprint {http://arxiv.org/abs/1704.00361} {arXiv:1704.00361 [gr-qc]} \BibitemShut {NoStop}%
\bibitem [{\citenamefont {Konoplya}\ \emph {et~al.}(2019{\natexlab{c}})\citenamefont {Konoplya}, \citenamefont {Zhidenko},\ and\ \citenamefont {Zinhailo}}]{Konoplya:2019hlu}%
  \BibitemOpen
  \bibfield  {author} {\bibinfo {author} {\bibfnamefont {R.~A.}\ \bibnamefont {Konoplya}}, \bibinfo {author} {\bibfnamefont {A.}~\bibnamefont {Zhidenko}}, \ and\ \bibinfo {author} {\bibfnamefont {A.~F.}\ \bibnamefont {Zinhailo}},\ }\href {\doibase 10.1088/1361-6382/ab2e25} {\bibfield  {journal} {\bibinfo  {journal} {Class. Quant. Grav.}\ }\textbf {\bibinfo {volume} {36}},\ \bibinfo {pages} {155002} (\bibinfo {year} {2019}{\natexlab{c}})},\ \Eprint {http://arxiv.org/abs/1904.10333} {arXiv:1904.10333 [gr-qc]} \BibitemShut {NoStop}%
\bibitem [{\citenamefont {Lütfüoğlu}(2025{\natexlab{d}})}]{Lutfuoglu:2025ljm}%
  \BibitemOpen
  \bibfield  {author} {\bibinfo {author} {\bibfnamefont {B.~C.}\ \bibnamefont {Lütfüoğlu}},\ }\href {\doibase 10.1140/epjc/s10052-025-14380-x} {\bibfield  {journal} {\bibinfo  {journal} {Eur. Phys. J. C}\ }\textbf {\bibinfo {volume} {85}},\ \bibinfo {pages} {630} (\bibinfo {year} {2025}{\natexlab{d}})},\ \Eprint {http://arxiv.org/abs/2504.18482} {arXiv:2504.18482 [gr-qc]} \BibitemShut {NoStop}%
\bibitem [{\citenamefont {Konoplya}\ \emph {et~al.}(2023)\citenamefont {Konoplya}, \citenamefont {Ovchinnikov},\ and\ \citenamefont {Ahmedov}}]{Konoplya:2023ahd}%
  \BibitemOpen
  \bibfield  {author} {\bibinfo {author} {\bibfnamefont {R.~A.}\ \bibnamefont {Konoplya}}, \bibinfo {author} {\bibfnamefont {D.}~\bibnamefont {Ovchinnikov}}, \ and\ \bibinfo {author} {\bibfnamefont {B.}~\bibnamefont {Ahmedov}},\ }\href {\doibase 10.1103/PhysRevD.108.104054} {\bibfield  {journal} {\bibinfo  {journal} {Phys. Rev. D}\ }\textbf {\bibinfo {volume} {108}},\ \bibinfo {pages} {104054} (\bibinfo {year} {2023})},\ \Eprint {http://arxiv.org/abs/2307.10801} {arXiv:2307.10801 [gr-qc]} \BibitemShut {NoStop}%
\bibitem [{\citenamefont {Dubinsky}(2024)}]{Dubinsky:2024aeu}%
  \BibitemOpen
  \bibfield  {author} {\bibinfo {author} {\bibfnamefont {A.}~\bibnamefont {Dubinsky}},\ }\href {\doibase 10.1016/j.dark.2024.101657} {\bibfield  {journal} {\bibinfo  {journal} {Phys. Dark Univ.}\ }\textbf {\bibinfo {volume} {46}},\ \bibinfo {pages} {101657} (\bibinfo {year} {2024})},\ \Eprint {http://arxiv.org/abs/2405.08262} {arXiv:2405.08262 [gr-qc]} \BibitemShut {NoStop}%
\bibitem [{\citenamefont {Dubinsky}\ and\ \citenamefont {Zinhailo}(2024)}]{Dubinsky:2024hmn}%
  \BibitemOpen
  \bibfield  {author} {\bibinfo {author} {\bibfnamefont {A.}~\bibnamefont {Dubinsky}}\ and\ \bibinfo {author} {\bibfnamefont {A.}~\bibnamefont {Zinhailo}},\ }\href {\doibase 10.1140/epjc/s10052-024-13206-6} {\bibfield  {journal} {\bibinfo  {journal} {Eur. Phys. J. C}\ }\textbf {\bibinfo {volume} {84}},\ \bibinfo {pages} {847} (\bibinfo {year} {2024})},\ \Eprint {http://arxiv.org/abs/2404.01834} {arXiv:2404.01834 [gr-qc]} \BibitemShut {NoStop}%
\bibitem [{\citenamefont {Dubinsky}(2025)}]{Dubinsky:2025fwv}%
  \BibitemOpen
  \bibfield  {author} {\bibinfo {author} {\bibfnamefont {A.}~\bibnamefont {Dubinsky}},\ }\href@noop {} {\bibfield  {journal} {\bibinfo  {journal} {International Journal of Gravitation and Theoretical Physics}\ }\textbf {\bibinfo {volume} {1}},\ \bibinfo {pages} {2} (\bibinfo {year} {2025})},\ \Eprint {http://arxiv.org/abs/2507.00256} {arXiv:2507.00256 [gr-qc]} \BibitemShut {NoStop}%
\bibitem [{\citenamefont {Konoplya}\ and\ \citenamefont {Abdalla}(2005)}]{Konoplya:2005sy}%
  \BibitemOpen
  \bibfield  {author} {\bibinfo {author} {\bibfnamefont {R.~A.}\ \bibnamefont {Konoplya}}\ and\ \bibinfo {author} {\bibfnamefont {E.}~\bibnamefont {Abdalla}},\ }\href {\doibase 10.1103/PhysRevD.71.084015} {\bibfield  {journal} {\bibinfo  {journal} {Phys. Rev. D}\ }\textbf {\bibinfo {volume} {71}},\ \bibinfo {pages} {084015} (\bibinfo {year} {2005})},\ \Eprint {http://arxiv.org/abs/hep-th/0503029} {arXiv:hep-th/0503029} \BibitemShut {NoStop}%
\bibitem [{\citenamefont {Bolokhov}\ and\ \citenamefont {Skvortsova}(2025{\natexlab{b}})}]{Bolokhov:2025lnt}%
  \BibitemOpen
  \bibfield  {author} {\bibinfo {author} {\bibfnamefont {S.~V.}\ \bibnamefont {Bolokhov}}\ and\ \bibinfo {author} {\bibfnamefont {M.}~\bibnamefont {Skvortsova}},\ }\href@noop {} {\bibfield  {journal} {\bibinfo  {journal} {International Journal of Gravitation and Theoretical Physics}\ }\textbf {\bibinfo {volume} {1}},\ \bibinfo {pages} {3} (\bibinfo {year} {2025}{\natexlab{b}})},\ \Eprint {http://arxiv.org/abs/2507.07196} {arXiv:2507.07196 [gr-qc]} \BibitemShut {NoStop}%
\bibitem [{\citenamefont {Bolokhov}(2024{\natexlab{b}})}]{Bolokhov:2024ixe}%
  \BibitemOpen
  \bibfield  {author} {\bibinfo {author} {\bibfnamefont {S.~V.}\ \bibnamefont {Bolokhov}},\ }\href {\doibase 10.1140/epjc/s10052-024-12990-5} {\bibfield  {journal} {\bibinfo  {journal} {Eur. Phys. J. C}\ }\textbf {\bibinfo {volume} {84}},\ \bibinfo {pages} {634} (\bibinfo {year} {2024}{\natexlab{b}})},\ \Eprint {http://arxiv.org/abs/2404.09364} {arXiv:2404.09364 [gr-qc]} \BibitemShut {NoStop}%
\bibitem [{\citenamefont {Malik}(2025{\natexlab{b}})}]{Malik:2024nhy}%
  \BibitemOpen
  \bibfield  {author} {\bibinfo {author} {\bibfnamefont {Z.}~\bibnamefont {Malik}},\ }\href {\doibase 10.1016/j.aop.2025.170046} {\bibfield  {journal} {\bibinfo  {journal} {Annals Phys.}\ }\textbf {\bibinfo {volume} {479}},\ \bibinfo {pages} {170046} (\bibinfo {year} {2025}{\natexlab{b}})},\ \Eprint {http://arxiv.org/abs/2409.01561} {arXiv:2409.01561 [gr-qc]} \BibitemShut {NoStop}%
\bibitem [{\citenamefont {Konoplya}\ and\ \citenamefont {Stashko}(2025)}]{Konoplya:2024lch}%
  \BibitemOpen
  \bibfield  {author} {\bibinfo {author} {\bibfnamefont {R.~A.}\ \bibnamefont {Konoplya}}\ and\ \bibinfo {author} {\bibfnamefont {O.~S.}\ \bibnamefont {Stashko}},\ }\href {\doibase 10.1103/PhysRevD.111.104055} {\bibfield  {journal} {\bibinfo  {journal} {Phys. Rev. D}\ }\textbf {\bibinfo {volume} {111}},\ \bibinfo {pages} {104055} (\bibinfo {year} {2025})},\ \Eprint {http://arxiv.org/abs/2408.02578} {arXiv:2408.02578 [gr-qc]} \BibitemShut {NoStop}%
\bibitem [{\citenamefont {Skvortsova}(2024{\natexlab{c}})}]{Skvortsova:2024wly}%
  \BibitemOpen
  \bibfield  {author} {\bibinfo {author} {\bibfnamefont {M.}~\bibnamefont {Skvortsova}},\ }\href {\doibase 10.1134/S020228932470018X} {\bibfield  {journal} {\bibinfo  {journal} {Grav. Cosmol.}\ }\textbf {\bibinfo {volume} {30}},\ \bibinfo {pages} {279} (\bibinfo {year} {2024}{\natexlab{c}})},\ \Eprint {http://arxiv.org/abs/2405.15807} {arXiv:2405.15807 [gr-qc]} \BibitemShut {NoStop}%
\bibitem [{\citenamefont {Leaver}(1985)}]{Leaver:1985ax}%
  \BibitemOpen
  \bibfield  {author} {\bibinfo {author} {\bibfnamefont {E.~W.}\ \bibnamefont {Leaver}},\ }\href {\doibase 10.1098/rspa.1985.0119} {\bibfield  {journal} {\bibinfo  {journal} {Proc. Roy. Soc. Lond. A}\ }\textbf {\bibinfo {volume} {402}},\ \bibinfo {pages} {285} (\bibinfo {year} {1985})}\BibitemShut {NoStop}%
\bibitem [{\citenamefont {Nollert}(1993)}]{Nollert:1993zz}%
  \BibitemOpen
  \bibfield  {author} {\bibinfo {author} {\bibfnamefont {H.-P.}\ \bibnamefont {Nollert}},\ }\href {\doibase 10.1103/PhysRevD.47.5253} {\bibfield  {journal} {\bibinfo  {journal} {Phys. Rev. D}\ }\textbf {\bibinfo {volume} {47}},\ \bibinfo {pages} {5253} (\bibinfo {year} {1993})}\BibitemShut {NoStop}%
\bibitem [{\citenamefont {Konoplya}\ and\ \citenamefont {Zhidenko}(2004)}]{Konoplya:2004uk}%
  \BibitemOpen
  \bibfield  {author} {\bibinfo {author} {\bibfnamefont {R.~A.}\ \bibnamefont {Konoplya}}\ and\ \bibinfo {author} {\bibfnamefont {A.}~\bibnamefont {Zhidenko}},\ }\href {\doibase 10.1088/1126-6708/2004/06/037} {\bibfield  {journal} {\bibinfo  {journal} {JHEP}\ }\textbf {\bibinfo {volume} {06}},\ \bibinfo {pages} {037} (\bibinfo {year} {2004})},\ \Eprint {http://arxiv.org/abs/hep-th/0402080} {arXiv:hep-th/0402080} \BibitemShut {NoStop}%
\bibitem [{\citenamefont {Dias}\ \emph {et~al.}(2022)\citenamefont {Dias}, \citenamefont {Godazgar},\ and\ \citenamefont {Santos}}]{Dias:2022oqm}%
  \BibitemOpen
  \bibfield  {author} {\bibinfo {author} {\bibfnamefont {O.~J.~C.}\ \bibnamefont {Dias}}, \bibinfo {author} {\bibfnamefont {M.}~\bibnamefont {Godazgar}}, \ and\ \bibinfo {author} {\bibfnamefont {J.~E.}\ \bibnamefont {Santos}},\ }\href {\doibase 10.1007/JHEP07(2022)076} {\bibfield  {journal} {\bibinfo  {journal} {JHEP}\ }\textbf {\bibinfo {volume} {07}},\ \bibinfo {pages} {076} (\bibinfo {year} {2022})},\ \Eprint {http://arxiv.org/abs/2205.13072} {arXiv:2205.13072 [gr-qc]} \BibitemShut {NoStop}%
\bibitem [{\citenamefont {Stuchl{\'\i}k}\ and\ \citenamefont {Zhidenko}(2025{\natexlab{a}})}]{Stuchlik:2025mjj}%
  \BibitemOpen
  \bibfield  {author} {\bibinfo {author} {\bibfnamefont {Z.}~\bibnamefont {Stuchl{\'\i}k}}\ and\ \bibinfo {author} {\bibfnamefont {A.}~\bibnamefont {Zhidenko}},\ }\href@noop {} {\  (\bibinfo {year} {2025}{\natexlab{a}})},\ \Eprint {http://arxiv.org/abs/2506.09829} {arXiv:2506.09829 [gr-qc]} \BibitemShut {NoStop}%
\bibitem [{\citenamefont {Xiong}\ and\ \citenamefont {Li}(2023)}]{Xiong:2023usm}%
  \BibitemOpen
  \bibfield  {author} {\bibinfo {author} {\bibfnamefont {W.}~\bibnamefont {Xiong}}\ and\ \bibinfo {author} {\bibfnamefont {P.-C.}\ \bibnamefont {Li}},\ }\href {\doibase 10.1103/PhysRevD.108.044064} {\bibfield  {journal} {\bibinfo  {journal} {Phys. Rev. D}\ }\textbf {\bibinfo {volume} {108}},\ \bibinfo {pages} {044064} (\bibinfo {year} {2023})},\ \Eprint {http://arxiv.org/abs/2305.04040} {arXiv:2305.04040 [gr-qc]} \BibitemShut {NoStop}%
\bibitem [{\citenamefont {Kanti}\ \emph {et~al.}(2006)\citenamefont {Kanti}, \citenamefont {Konoplya},\ and\ \citenamefont {Zhidenko}}]{Kanti:2006ua}%
  \BibitemOpen
  \bibfield  {author} {\bibinfo {author} {\bibfnamefont {P.}~\bibnamefont {Kanti}}, \bibinfo {author} {\bibfnamefont {R.~A.}\ \bibnamefont {Konoplya}}, \ and\ \bibinfo {author} {\bibfnamefont {A.}~\bibnamefont {Zhidenko}},\ }\href {\doibase 10.1103/PhysRevD.74.064008} {\bibfield  {journal} {\bibinfo  {journal} {Phys. Rev. D}\ }\textbf {\bibinfo {volume} {74}},\ \bibinfo {pages} {064008} (\bibinfo {year} {2006})},\ \Eprint {http://arxiv.org/abs/gr-qc/0607048} {arXiv:gr-qc/0607048} \BibitemShut {NoStop}%
\bibitem [{\citenamefont {Zinhailo}(2024{\natexlab{b}})}]{Zinhailo:2024kbq}%
  \BibitemOpen
  \bibfield  {author} {\bibinfo {author} {\bibfnamefont {A.~F.}\ \bibnamefont {Zinhailo}},\ }\href {\doibase 10.13140/RG.2.2.26785.01124} {\  (\bibinfo {year} {2024}{\natexlab{b}}),\ 10.13140/RG.2.2.26785.01124}\BibitemShut {NoStop}%
\bibitem [{\citenamefont {Konoplya}\ and\ \citenamefont {Zhidenko}(2007)}]{Konoplya:2007zx}%
  \BibitemOpen
  \bibfield  {author} {\bibinfo {author} {\bibfnamefont {R.~A.}\ \bibnamefont {Konoplya}}\ and\ \bibinfo {author} {\bibfnamefont {A.}~\bibnamefont {Zhidenko}},\ }\href {\doibase 10.1103/PhysRevD.76.084018} {\bibfield  {journal} {\bibinfo  {journal} {Phys. Rev. D}\ }\textbf {\bibinfo {volume} {76}},\ \bibinfo {pages} {084018} (\bibinfo {year} {2007})},\ \bibinfo {note} {[Erratum: Phys.Rev.D 90, 029901 (2014)]},\ \Eprint {http://arxiv.org/abs/0707.1890} {arXiv:0707.1890 [hep-th]} \BibitemShut {NoStop}%
\bibitem [{\citenamefont {Stuchl{\'\i}k}\ and\ \citenamefont {Zhidenko}(2025{\natexlab{b}})}]{Stuchlik:2025ezz}%
  \BibitemOpen
  \bibfield  {author} {\bibinfo {author} {\bibfnamefont {Z.}~\bibnamefont {Stuchl{\'\i}k}}\ and\ \bibinfo {author} {\bibfnamefont {A.}~\bibnamefont {Zhidenko}},\ }\href {\doibase 10.1103/qv83-1jw3} {\bibfield  {journal} {\bibinfo  {journal} {Phys. Rev. D}\ }\textbf {\bibinfo {volume} {112}},\ \bibinfo {pages} {024064} (\bibinfo {year} {2025}{\natexlab{b}})},\ \Eprint {http://arxiv.org/abs/2503.06775} {arXiv:2503.06775 [gr-qc]} \BibitemShut {NoStop}%
\bibitem [{\citenamefont {Bolokhov}(2024{\natexlab{c}})}]{Bolokhov:2023ruj}%
  \BibitemOpen
  \bibfield  {author} {\bibinfo {author} {\bibfnamefont {S.~V.}\ \bibnamefont {Bolokhov}},\ }\href {\doibase 10.1103/PhysRevD.109.064017} {\bibfield  {journal} {\bibinfo  {journal} {Phys. Rev. D}\ }\textbf {\bibinfo {volume} {109}},\ \bibinfo {pages} {064017} (\bibinfo {year} {2024}{\natexlab{c}})}\BibitemShut {NoStop}%
\end{thebibliography}%

\end{document}